# A dynamic intermediate state limits the folding rate of a discontinuous two-domain protein

Ganesh Agam[1], Anders Barth[1,†,] and Don C. Lamb[1]*

[1] Department of Chemistry, Center for NanoScience, Nanosystems Initiative Munich (NIM), and Center for Integrated Protein Science Munich (CiPSM), Ludwig-Maximilians University Munich, Munich, Germany.

[†] Current address: Department of Bionanoscience, Kavli Institute of Nanoscience Delft, Delft University of Technology, 2629, HZ, Delft, The Netherlands.

* **corresponding author: d.lamb@lmu.de**

## ABSTRACT

Protein folding is an indispensable process for the majority of proteins after their synthesis from ribosomes in the cell. Most *in vitro* protein folding studies have focused on small, single-domain proteins, which account for approximately one-third of all proteins. Hence, it is also important to understand the folding process of large, multi-domain proteins, especially when the domains are discontinuous. To study the co-dependent folding of two globular discontinuous domains, we choose the Maltose binding protein (MBP) as a model system. In particular, we studied a mutant of MBP that folds slowly and contains two mutations in the N-terminal domain. Despite the wide use of the double mutant MBP (DM-MBP) in studies of the cellular folding machinery, a systematic study of its domain-wise folding is still missing. Here, using two- and three-color single-molecule Förster resonance energy transfer (smFRET) experiments, we study the refolding of both the domains and the interaction between the domains of DM-MBP. Initial two-color smFRET measurements of the N-terminal domain (NTD) reveal the presence of a folding intermediate, also known from previous studies. The same folding intermediate is observed in measurements monitoring the C- terminal domain (CTD) and the NTD-CTD (N-C) interface. The refolding intermediate is not due to the aggregation in the refolding reaction and is dynamic on the sub-millisecond timescale. The dynamics governs DM-MBP folding by shifting the equilibrium between native and unfolded conformation depending on the denaturant concentrations. Quantitative analysis on underlying dynamic interconversions revealed a delay in NTD folding imposed by the entropic barrier being the primary cause for slow DM-MBP folding. Moreover, highly dynamic CTD folds after NTD completes the folding. Using three-color smFRET, we could show the NTD folds first and CTD later in a same protein. Molecular dynamic simulations for temperature-induced unfolding on WT- and DM-MBP identify a folding nucleus in the NTD, which is rich in hydrophobic residues, and explains why the two mutations in this folding nucleus slow down the folding kinetics. In the presence of the bacterial Hsp60 chaperonin system GroEL/ES, we observe that DM-MBP is still dynamic within the chaperonin cavity but the chaperonin limits the conformational space that the substrate explores. Hence, confinement aids DM-MBP in overcoming the entropically limited folding barrier. The study reports on the subtle tuning and co-dependency for protein folding between two-domains with a discontinuous arrangement.

**INTRODUCTION**

The three-dimensional structure of a protein is defined by its primary amino acid sequence (Anfinsen, 1973). After its synthesis from the ribosome as a polypeptide chain, most of the proteins, with the exception of intrinsically disordered proteins, fold into their native structure either spontaneously or with the help of chaperones. Though the ultimate aim is to understand the folding process *in vivo* and studies of folding on ribosomes are becoming technically possible (Holtkamp et al., 2015)*, in vitro* unfolding-refolding studies allow precise control of the folding/unfolding conditions and investigation of the process in isolation (Sela, White, & Anfinsen, 1957). Most folding studies have focused on simple, small proteins (upper limit of ~100 amino acids) with a few exceptions (Batey, Nickson, & Clarke, 2008). However, it is estimated that ~67% of the eukaryotic proteins have more than a single domain (Teichmann, Parkt, & Chothia, 1998) (Chothia, Gough, Vogel, & Teichmann, 2003)(Han, Batey, Nickson, Teichmann, & Clarke, 2007), with the majority being two-domain proteins (Jones et al., 1998). Two-domain proteins typically show continuous domain topology, ensuring the efficient folding of each domain independently. Continuous-domain proteins comprise about ~72% of all two-domain proteins (Jones et al., 1998). The remaining ~28% of two-domain proteins have discontinuous domains with the insertion of at least one region from one domain within the other. Discontinuous domains make the folding process energetically inefficient as the domains are co-dependent for assuming the final fold (Wetlaufer, 1973), (Collinet et al., 2000). Most enzymes and regulatory proteins exhibit this type of complex topology (Shank, Cecconi, Dill, Marqusee, & Bustamante, 2010) (Shank 2010). Despite their abundance in catalysis and in allosteric processes, little is known about their folding process. Most of the small proteins generally fold spontaneously within a short time (microseconds to milliseconds) (Chung, McHale, Louis, & Eaton, 2012). Different theories have been put forward to explain the process of folding (Wolynes, Onuchic, & Thirumalai, 1995) (Thirumalai, O'Brien, Morrison, & Hyeon, 2010). Large proteins having multiple domains fold relatively slow (seconds to minutes) and exhibit a complex multi-phase folding process (Arai, Iwakura, Matthews, & Bilsel, 2011), (Wolynes et al., 1995), (Onuchic & Wolynes, 2004). In the case of two-domain proteins, especially with a discontinuous domain topology, the main questions are whether the domains fold sequentially or simultaneously in a cooperative way.

Here, we investigated folding of the Maltose binding protein (MBP). MBP is a monomeric ~42 kDa protein having two discontinuous domains, namely the N- and C-terminal domains (NTD and CTD). Both domains contain $\alpha$-$\beta$-$\alpha$ secondary structural motifs although the $\beta$-sheets extend over both the domains. The maltose-binding pocket is positioned between the domains (Spurlino, Lu, & Quiocho, 1991). The folding of WT-MBP and several folding-defective mutants have been studied previously (Chun, Strobel, Bassford, & Randall, 1993). One of them, containing the mutations V8G and Y283D in the NTD, folds with a significantly slower rate (half-life, $t_{1/2}$ ~30 min) as compared to WT-MBP ($t_{1/2}$ ~25 s) (Tang et al., 2006). Interestingly, this double-mutant MBP (DM-MBP) shows intermediate states during the folding process (Sharma et al., 2008), (Chakraborty et al., 2010). The structure and domain-wise conformation of these intermediates is currently unknown but would be of high interest due to the discontinuous domain arrangement of DM-MBP (Wetlaufer, 1973) (Collinet et al., 2000).

Single-molecule Förster resonance energy transfer (smFRET) methods have been instrumental in understanding the conformational heterogeneity and dynamic properties of biomolecules. In two-color smFRET, one fluorophore (donor) transfers its energy non-radiatively to another fluorophore (acceptor), whereby the efficiency of energy transfer strongly depends on the distance between them in the range of 2-10 nm (Forster, 1946) (Ha et al., 1996). To avoid artifacts due to surface immobilization, most smFRET-based folding studies have been performed on freely diffusing molecules except small domain proteins (Deniz et al., 2000), (Schuler, Lipman, & Eaton, 2002), (Chung et al., 2012). Folding is the largest conformational change a protein can undergo. It is thought to be a cooperative process involving all parts of the protein (Wolynes et al., 1995). Compared to ensemble folding experiments, smFRET can investigate the heterogeneity of the populations present within the folding process (Borgia et al., 2011). SmFRET folding studies have led to insights into spontaneous folding, folding intermediates, and the regulation of folding pathways by various chaperone systems (Schuler et al., 2002), (Sharma et al., 2008), (Hofmann et al., 2010), (Kellner et al., 2014) (Dahiya et al., 2019). However, it remains challenging to study the folding pathway of a large protein with multiple domains by two-color smFRET (Pirchi et al., 2011). Multi-color smFRET has the unique advantage of monitoring multiple distances simultaneously (Hohng, Joo, & Ha, 2004), (Clamme & Deniz, 2005), (Nam et al., 2007). By positioning multiple fluorophores specifically on each domain, one can study the folding of individual domains in the context of the folding pathway of the whole protein. Recent advancements in diffusion- and surface-based multi-color smFRET single-molecule experiments have been applied to unravel the inter or intra-molecular correlated motions in biomolecules, (Gambin & Deniz, 2010), (Ratzke, Hellenkamp, & Hugel, 2014), (Barth, Voith Von Voithenberg, & Lamb, 2019), (Voith von Voithenberg et al., 2021).

In the current study, we characterize the known folding intermediate of DM-MBP and its importance in the overall folding. Furthermore, we ask the question whether both DM-MBP domains fold simultaneously or in a particular order. We address this by performing diffusion-based two- and three-color smFRET measurements of the NTD, CTD and of the N-C interface of DM-MBP. We characterized the previously reported intermediate state in the folding of DM-MBP under non-aggregating conditions (Chakraborty et al., 2010). Separate two-color smFRET measurements of the NTD, CTD and the N-C interface at different denaturant conditions revealed the unique conformation of the intermediate that is present in all the three probed coordinates. Moreover, plots of the FRET efficiency versus donor fluorescence lifetime show that the intermediate state is dynamic on the sub-millisecond timescale and fluctuates between native and unfolded conformations of the protein. Quantitative information on the microscopic rates of the conformational transitions governing the intermediate state were extracted employing a dynamic photon distribution analysis (PDA). Comparing the macroscopic and the microscopic rates of refolding confirmed that the intermediate state limits the folding mainly due to an entropic barrier. Analysis of the individual domains illustrated the delayed folding in the NTD being the primary cause for overall slow folding imposed by the double mutations. As soon as the NTD folds, folding of highly dynamic CTD follows. Three-color smFRET allows to address the coordination of the conformations for three different positions at the same time. For three-color smFRET, the labeling positions were chosen such that the three inter-dye distances report on the folding of the NTD, CTD and N-C domain interface. Equilibrium three-color smFRET measurements provided evidence that the intermediate state is present in all the three coordinates simultaneously. Furthermore, probing three distances simultaneously shows that the native state of the two domains is assumed in a sequential way where first NTD folds which was followed by CTD , consistent with the results from the dynamic PDA on individual domains. Using molecular dynamics (MD) simulations, we investigate the temperature-induced unfolding of WT and DM-MBP and provide insights into the effect of the point mutations on the folding pathway. MD simulations are consistent with the finding that the DM-MBP folding has to cross an entropic barrier, which delays folding. We also investigated the effect of GroEL\ES chaperonin, a bacterial Hsp60 chaperone and Hsp10 co-chaperone system, on the DM-MBP folding. It was previously shown that GroEL\ES accelerates the DM-MBP folding by 8-13 fold. We found that DM-MBP is still dynamic when bound to GroEL and when encapsulated within the cavity of GroEL\ES. Interestingly, when bound to GroEL alone, the equilibrium of the dynamics of DM-MBP is slightly shifted towards the unfolded conformation whereas, in co-operation with GroES, cavity confinement leads to a compaction of the dynamic folding intermediate state towards the native state. Hence, by limiting the conformational space explored by DM-MBP, the chaperonin helps DM-MBP to overcome the entropic barrier during folding. The proposed mechanism suggests one pathway for the folding of discontinuous multi-domains on physiological timescales when the folding of the individual domains is interdependent and how chaperonins accelerate their folding by modulating the folding landscape.

## RESULTS

**Characterization of a previously known hysteresis in the folding of DM-MBP**

We first characterized the denaturation/refolding of DM-MBP using intrinsic tryptophan fluorescence at the ensemble level, taking advantage of the eight tryptophan residues present in DM-MBP (Figure 1A) (PDB ID:1OMP) (Sharff, Rodseth, Spurlino, & Quiocho, 1992). Burial of solvent exposed hydrophobic tryptophan residues inside the protein core upon folding results in an increase in the tryptophan fluorescence. For comparison, we also performed the denaturation/refolding of WT-MBP. When MBP is completely unfolded in 3 M guanidine-hydrochloride (GuHCl), the tryptophan fluorescence is reduced by ~2.5-fold as compared to the native protein. We first performed equilibrium measurements of unfolding and refolding by titrating with different GuHCl concentrations. The MBP concentration for these experiments was kept at 40 nM, lower than the previous studies to avoid aggregation (Figure 1B). In the case of unfolding, when native MBP was denatured with increasing concentrations of denaturant, both WT and DM-MBP starts to unfold above 0.6 M GuHCl. Complete unfolding was observed above 1 M GuHCl. Strikingly, compared to WT which follows the similar trend during refolding as denaturation, DM-MBP only refolds from the denatured state starting at a concentration of 0.4 M GuHCl. The hysteresis in formation of the native state with respect to unfolding under similar denaturing conditions has been attributed to kinetic traps present in the folding landscape of the protein (Andrews, Capraro, Sulkowska, Onuchic, & Jennings, 2013). The same observation was made by Chakraborty et al. 2010, who also proposed a kinetically trapped state in DM-MBP. Notably, DM-MBP completely refolds to its native state below a final concentration of 0.1 M GuHCl in refolding buffer (Figure 1B). Hence, for all the ensemble and single molecule refolding kinetic measurements performed throughout this study, we kept the final GuHCl concentrations below 0.1 M. Next, we measured the kinetics of refolding. Denatured WT and DM-MBP in 3 M GuHCl was allowed to refold after a 75-fold dilution at a final concentration of 40 nM MBP and 40 mM GuHCl. The refolding half-time ($t_{1/2}$) for WT and DM-MBP was found to be 0.29±0.07 and 22.70±2.65 min with a refolding rate of 2.44±0.53 and 0.028±0.003 min$^{-1}$ respectively (Figure 1C) (Table 1).

The delayed refolding giving rise to the hysteresis effect might be caused by inter-molecular interactions (i.e. aggregation) of the unfolded peptide chain due to the high degree of exposed hydrophobic residues. To exclude this possibility, we used nanomolar concentrations of protein for the tryptophan fluorescence assays. To verify that there is no aggregation at the nanomolar concentrations used for these measurements, we performed fluorescence cross-correlation spectroscopy (FCCS) experiments. FCCS, in combination with pulsed interleaved excitation (PIE) (Müller, Zaychikov, Bräuchle, & Lamb, 2005) is highly sensitive to the presence of any interaction between molecules labeled with different fluorophores, giving rise to a cross-correlation signal. As a positive control, we measured double-stranded DNA labeled with both Atto532 and Atto647N, which gives rise to a significant cross-correlation signal (red curve in Figure 1D). A single-cysteine mutant of DM-MBP (A52C) labeled with either Atto532 or Alexa647 sample (500 - 1000 nM) were unfolded in 3 M GuHCl for 30 min at 50°C, mixed in equal amounts and diluted to allow refolding in 0.1 M GuHCl at a final protein concentration of 40 nM, similar to the trp florescence experiments. No cross-correlation amplitude was detected, indicating that no aggregation occurred in the refolding over the course of 60 minutes (green, magenta and yellow curves in Figure 1D). As a negative control, a mixture of freely-diffusing dyes was also measured and showed no cross-correlation signal (blue curve in Figure 1D). The observed hysteresis effect and refolding rate of DM-MBP are in good agreement with previous reports (Chakraborty et al., 2010) (Tang et al., 2006). In addition, we could verify that the previously observed hysteresis is not induced by aggregation.

**smFRET identifies a unique intermediate population as a kinetically trapped state**

To gain further insights into the structural properties of the trapped state of DM-MBP during refolding, we performed single molecule two-color smFRET measurements using multiparameter fluorescence combined with PIE (MFD-PIE) (Kudryavtsev et al., 2012). By using picomolar (pM) concentrations of labeled molecules in combination with the small observation volume (~fL) of a confocal microscope, bursts of fluorescence signal are observed for single molecules diffusing through the observation spot on the order of several milliseconds. Using MFD-PIE, the fluorescence signal obtained for single molecule events contains information about the labelling stoichiometry, FRET efficiency, fluorescence lifetime and anisotropy. The type of information that can be collected is shown in Supplementary Figure 1 (see Materials and Methods). Since the two folding mutations are in the NTD of MBP, we first investigated the refolding of the NTD. To probe the conformation of the NTD, a double-cysteine mutant of DM-MBP (52C-298C) was stochastically labeled with the fluorophores Atto532 (donor) and Alexa647 (acceptor) using cysteine-maleimide chemistry (Figure 1A) (Materials and Methods). The Förster distance, the inter-dye separation at which the FRET efficiency reduces to 50%, was determined to be ~62 Å for Atto532 and Alexa647 (Voith von Voithenberg et al., 2016) (Table 2). The two-color smFRET measurements exhibited a FRET efficiency of ~0.85 for the native NTD (Figure 2A, first row of FRET efficiency histograms). From a photon distribution analysis (PDA) of the smFRET efficiency histogram, we calculated an inter-dye distance distribution with an average separation of 48.3 Å. From the crystal structure (PDB ID:1OMP), an accessible volume (AV) calculation for the given dyes yielded a distance of 51.3 Å (Table 3) (Supplementary Figure 2A) (Materials and Methods) (Antonik, Felekyan, Gaiduk, & Seidel, 2006) (Kalinin et al., 2012). Thus, the experimentally determined distance is indicative of a folded NTD. When we measured the NTD conformation under denaturing conditions in 2 M GuHCl, a FRET efficiency of ~0.1 was obtained (Figure 2A, last row of FRET efficiency histograms). For this unfolded state, the PDA analysis yielded an average separation of ~83.8 Å with a broad width (17 Å) (Table 3). The above FRET efficiency values and donor-acceptor separations for both the native and unfolded state agrees well with previous smFRET measurements using fluorophores with similar spectroscopic properties attached at the same labeling positions (Sharma et al., 2008).

Next, we performed equilibrium unfolding and refolding measurements on the NTD at different GuHCl concentrations (Figure 2A), similar to the unfolding and refolding measurements performed using tryptophan fluorescence (Figure 1C). Gaussian fits to the obtained FRET efficiency distributions identified one intermediate FRET population with ~0.6 FRET efficiency, in addition to the native and completely unfolded states. Note that, as the denaturant concentration decreases under the denaturing conditions from 2 M to 0.9 M, the FRET efficiency of the unfolded state changes from ~0.1 to ~0.3 due to compaction of the denatured polypeptide at low GuHCl concentrations. To trace the origin of the intermediate population, the obtained average FRET efficiencies were compared with the tryptophan fluorescence measurements. Interestingly, during the refolding measurements, there is a significant contribution of the intermediate FRET population to the smFRET histograms for measurements collected within the hysteresis range of 0.4-1 M GuHCl (Figure 2B). The above analysis gives a glimpse into the folding process: starting from the completely unfolded state of ~0.08 FRET efficiency, there is a gradual hydrophobic collapse (depending on the GuHCl concentration in the buffer), and then the protein starts to refold into the native state. During refolding from the collapse state to the native state, a peak is observed in the smFRET histogram with an intermediate FRET efficiency of 0.6.

To analyze the refolding protein under non-equilibrium conditions, we performed a kinetic analysis of a smFRET measurement where refolding was initiated in denatured DM-MBP by a 75-fold dilution of the denaturant. Indeed, during the first 2000 s of the measurement, the refolding NTD had a FRET efficiency of 0.6 in the beginning and completely folds to a native state of 0.85 FRET efficiency (Figure 2C). From a quantitative analysis of the FRET efficiency histograms, we obtained a $t_{1/2}$ of 29.75±0.39 min, a similar rate as to what was obtained for refolding of an unlabeled protein (Figure 2D) (Table 1). This indicates that the fluorophores do not interfere with the refolding process of DM-MBP. In summary, the above equilibrium and kinetic smFRET experiments probing NTD has identified an intermediate state unique to the refolding reaction.

**The intermediate state is also present in the CTD and N-C interface**

Having characterized the intermediate state in the NTD, we investigated whether refolding of the CTD and formation of the N-C interface also exhibit the hysteresis. To monitor the CTD conformation, we made use of the previously characterized double-cysteine mutant of DM-MBP (175C-298C) (Figure 1A) (Sharma et al., 2008). Equilibrium unfolding and refolding curves were recorded using two-color smFRET measurements on the CTD labeled with the same dye-pair as for the NTD (Atto532 and Alexa647, Figure 3A). The native and completely unfolded states of the CTD have similar donor-acceptor separations as for the native and completely unfolded state of NTD, respectively (Supplementary Figure 2B) (Table 3). A plot of the FRET efficiency versus the titrated GuHCl concentration for folding and unfolding show a similar trend as for the NTD where unfolding behaves as a two-state system and refolding has a similar intermediate population with a FRET efficiency of 0.6 (Figure 3A). Kinetic experiments on CTD refolding yielded a $t_{1/2}$ of 19.28±7.88 min (Figure 3B). A similar observation was made for equilibrium and kinetic measurements on the N-C interface (52C-175C) (Figure 3C,D). For the N-C interface, the intermediate population is more compact with a FRET efficiency of 0.85. The structure of the N-C interface is well defined in the native state, having a small width of 2.8 Å for a inter-dye distance of 44.8 Å as determined by PDA. The AV simulations predict an inter-dye separation of 40.0 Å (Supplementary Figure 2C and Table 3). The unfolded state has a distance of 74.9 Å, comparable to the NTD and CTD unfolded states (Table 3). The kinetics of refolding are similar to the other mutants and to the unlabeled protein, which confirms the functionality of the both the labeled CTD and N-C interface constructs (Figure 3B and 3D). Due to the similar FRET efficiencies of the folded and the kinetically trapped state in the N-C interface construct, a kinetic analysis based on the FRET efficiency histograms is not possible. Hence, we made use of the observation that the fluorescence lifetime of Alexa647 is different in the unfolded and the refolded protein (Supplementary Figure 3A). Due to steric restriction, the cis-trans isomerization of Alexa647 is hindered in the unfolded state, typical for cyanine-based dyes. The dye hence exhibits a higher lifetime of ~1.7 ns for the unfolded state, but has a lifetime of ~1.2 ns for the folded protein, close to that of free dye in water (Supplementary Figure 3A-B and Table S1). Despite the different acceptor lifetimes, the measured anisotropy for Alexa647 remained below ~0.25 for all constructs (Table S3). To ensure that the observed intermediate was not an artifact arising from the acceptor, we measured refolding with a different acceptor, Atto647N, where the same intermediate state was measured (Supplementary Figure 3C). Appropriate corrections were made to account for differences in the acceptor quantum yield for the different populations for all constructs labeled with Alexa647 (Supplementary Figure 3D and Table 2).

The above smFRET measurements on the NTD, CTD and N-C interface all exhibit the presence of an intermediate population during folding. To explore the origin of this universal intermediate in DM-MBP folding, we carried out two-color smFRET experiments on WT-MBP folding on one of the domains, in this case the CTD. As WT-MBP folds within a minute, it should not have any intermediate population visible in the 0.4-1 M GuHCl refolding titrations. Indeed, both unfolding and refolding curves have the same transition between the unfolded and native state, without any visible intermediate population (Supplementary Figure 4). This confirms that the observed intermediate state is unique to DM-MBP and absent in the wildtype protein.

**The Intermediate state is a dynamically averaged state undergoing sub-millisecond dynamics**

Next, we investigated whether the intermediate state is structured or dynamic. Here, we made use of the fluorescence lifetime of the donor fluorophore, which is available in MFD-PIE experiments (Supplementary Figure 1). In the absence of dynamics, there is a linear relationship between the FRET efficiency and the donor fluorescence lifetime (static FRET-line, black line in upper panels of Figure 4A-C, Materials and Methods). When a molecule interconverts between different conformations (with different FRET efficiencies that correspond to a specific donor fluorescence lifetime) during the observation time, a deviation from the static FRET-line is observed (Materials and Methods) (Kalinin, Valeri, Antonik, Felekyan, & Seidel, 2010). We performed a lifetime analysis of two-color measurements for all the three constructs (Table 4, Figure 4A-C). In all cases, the intermediate states deviate from the static FRET-line, indicating the presence of conformational dynamics. The observed deviation can be described by a single dynamic-FRET line (red line in upper panels of Figure 4A-C) connecting the unfolded and native states. This indicates that the intermediate state is not a distinct structural state, but originates from dynamic switching between the unfolded and folded states during the observation time (~ms, Figure 4D). The intermediate state is thus a dynamically averaged state where conformational fluctuations of the molecule much faster than the burst duration lead to a species-averaged FRET efficiency in the intensity-based calculations. To determine the FRET efficiencies of the end conformations, we fit the donor fluorescence lifetime data from all detected bursts to a biexponential model function (Supplementary Figure 5). The determined FRET efficiencies were compared with the intensity averaged FRET efficiency values and are plotted as white boxes in the lower panels of Figure 4A-C (Table 4). Note that the equilibrium between the unfolded and folded states shifts with denaturant concentrations and, hence, a corresponding FRET efficiency averaging takes place. Moreover, changing the acceptor dye does not affect the dynamics present in the system (Supplementary Figure 6).

Interestingly, when comparing the refolding curves for all the three constructs, we observed that the refolding reaction performed in 0.1 M GuHCl where DM-MBP completely refolds, NTD and N-C interface significantly showed folded fraction whereas CTD still has substantial fraction in the intermediate state (lower panels of Figure 4A-C). We looked into more detail in this direction in the following section finding the kinetics of the underlying transitions.

**Dynamic Photon distribution analysis quantifies the folding order of the domains**

To quantify the microscopic rates for the conformational transitions during refolding, we employed dynamic PDA (Kalinin, Valeri, et al., 2010). This analysis routine quantifies the rates at which subpopulations interconvert during the burst duration of a few milliseconds. The distance distributions are modeled on the raw photon counts to quantify the broadening beyond shot-noise. The robustness of the fit is increased by generating FRET efficiency histograms (or proximity ratio histograms) of different time resolutions (or binning the photons) on the order of milliseconds (Material and Methods) and globally fitting the histograms. The distances, widths and kinetic rates can be extracted from the analysis to determine the microscopic rates between the states, where states were defined by donor lifetimes (Table 4).

Table 5 summarizes the microscopic rates for conformational fluctuations between the natively folded ($F$) and unfolded state ($U$). For all three constructs, the unfolding rate ($k_{F \to U}$) increases with increasing GuHCl concentration (Figure 5A-C) (Supplementary Figure 7). This is consistent with a loss of the natively stabilized contacts by the increasing amount of GuHCl in solution. In contrast, there are striking differences in the folding rates ($k_{U \to F}$) of the three constructs. For the CTD and N-C interface, the transition rates remained unchanged between 2-3 ms$^{-1}$ for the CTD and between 3.3 and 3.9 ms$^{-1}$ for the N-C interface throughout all the refolding curves (Supplementary Figure 7) with the exception of refolding of the N-C interface in 0.1 M GuHCl (Table 5). As GuHCl destabilizes native contacts, the refolding rates are expected to be independent of GuHCl concentration. The reduction in the refolding kinetics of the N-C interface at 0.1 GuHCl is most likely due to the slower structural rearrangements necessary during the final folding step. For the NTD, a more complicated pattern is observed. The unfolding rates ($k_{F \to U}$) are between 2-3 ms$^{-1}$ for lower GuHCl concentrations and then slow down to 0.6 - 0.9 ms$^{-1}$ at higher GuHCl concentrations. One possible explanation is that the hydrophobic core in the NTD defined by the residues 8-21, 22-24 and 45-63 is needed to stabilize the folded state (Ye, Mayne, Kan, & Englander, 2018). Disruption of the hydrophobic core by GuHCl also decreases the refolding transition monitored with the NTD FRET construct. The similar relaxation times for the NTD and the N-C interface indicates cooperativity in the folding process.

With the given dynamic PDA analysis, we conclude that the CTD folds and unfolds quickly, waiting for the NTD to fold to stabilize the structure. Once the NTD finds the correct structure, both the CTD and N-C interface can lock in place. The combined microscopic rates of 0.75-0.84 ms$^{-1}$ for the transitions in the NTD and N-C interface and 3.4 ms$^{-1}$ in CTD are much faster than the macroscopic folding rate of 0.028 min$^{-1}$ ($2.8 \times 10^{-7}$ ms$^{-1}$) for DM-MBP (Table 1). This provides direct evidence for the rate-limiting step in DM-MBP folding being the native-like contact formation. From Kramer's equation, the difference in Gibbs free energy ($\Delta G_f$) required to cross the folding barrier can be calculated directly from the folding times (Kramers, 1940), (Schuler et al., 2002),

$$\tau_f \approx 2\pi\tau_0 exp\left(\frac{\Delta G_f}{RT}\right) \qquad (1)$$

where $\tau_f$ is the folding time, $R$ is the gas constant and $T$ is the temperature. Here, an assumption is made for harmonic frequencies to be uniform at the surface of an unfolded well and the top of the barrier. $\tau_0$ is the reconfiguration time in the unfolded well and hence defines the pre-exponential factor. $\tau_0$ is assumed to be $\tau_0 \approx$ 100 ns based on what has been observed in other proteins (Soranno et al., 2012) (Soranno et al., 2017), the above assumption and from the folding time, $\tau_f \sim 2 \times 10^6$ ms (33.3 minutes) for DM-MBP folding, a $\Delta G_f$ of 39 kJ/mol is required to achieve folding. When compared to the WT-MBP case, with the folding rate from ensemble measurement of 1.8 min$^{-1}$ ($\tau_f \sim 33.3 \times 10^3$ ms), a Gibbs free energy for folding of 29.3 kJ/mol is observed (Figure 5D). The additional 9.7 kJ/mol Gibbs free energy difference in the DM-MBP folding is unlikely to be enthalpic in nature and could be explained by the extra dynamics due to the absence of the hydrophobic core in DM-MBP. To verify that the barrier in DM-MBP folding is entropic in nature, trimethylamine N-oxide (TMAO) was incorporated in the assay buffer when studying DM-MBP refolding. TMAO is known to act as a chemical chaperone that confines the configuration space available to the protein by stabilizing the solvent shell around the protein (Bandyopadhyay et al., 2012). The confinement of DM-MBP by TMAO, similar to the cavity of GroEL/ES (Chakraborty et al., 2010), accelerated the formation of the native-like structure as observed for refolding of the NTD in 0.2 M GuHCl, where DM-MBP has significant population of the flexibility driven-trapped intermediate state without TMAO (Figure 5E). These finding are consistent with the notion that the folding intermediate has indeed an entropic barrier rather the local kinetic trap as TMAO would stabilize the kinetic trap further.

**The native and unfolded state of DM-MBP can be probed with three-color FRET**

Two-color FRET allowed us to probe the folding of the different domains separately. To obtain a coherent global view of DM-MBP folding, we applied three-color smFRET. Multi-color smFRET has the potential to visualize the correlative motions in different parts of biomolecules in real time (Gambin & Deniz, 2010), (Ratzke et al., 2014),(Barth, Voith Von Voithenberg, & Lamb, 2019), (Voith von Voithenberg et al., 2021). Using three labels on the protein, we can probe the folding of the two domains and the inter-domain interface simultaneously using the three FRET efficiencies between the blue, green and red fluorophores (Figure 6A) (Supplementary Figure 8A). For three-color FRET, we used the dyes Atto488, Atto565 and Alexa647, which were chosen to maximize the use of visible spectrum with a distance sensitivity indicated by their respective Förster distances (~50-70 Å) (Table 2). From a three-color smFRET measurement, we obtain the three FRET efficiencies between the different dye pairs: blue and green (BG), blue and red (BR) and green and red (GR), allowing us to address the three intramolecular distances in one MFD-PIE smFRET experiment (Materials and Methods, Supplementary Figure 8B-E) (Barth, Voith Von Voithenberg, & Lamb, 2019). For the three-color smFRET experiment, it is important to label the three fluorophores site-specifically. To this end, we incorporated an unnatural amino acid at one position, while ensuring selective labeling of the two cysteines using their differential solvent accessibility when maltose is bound (Supplementary Figure 9) (Materials and Methods). The triple-labeled DM-MBP was confirmed to be able to fold by assaying the kinetics of Alexa647 lifetime change upon refolding, which was found to be similar to the unlabeled protein (Supplementary Figure 8F) (Table 1 and S2).

First, we measured the native, denatured and refolded state of DM-MBP. According to the labeling positions, the FRET efficiency between the blue and green dyes (BG) reports on the NTD conformation, while the FRET efficiency between the green and red dyes (GR) corresponds to the CTD conformation and the FRET efficiency between the blue and red dyes (BR) monitors the N-C interface (Figure A). The three-color smFRET measurement of triple-labeled DM-MBP in the native state showed a high FRET peak at ~0.9 for all the three FRET efficiencies. Refolded DM-MBP showed the same FRET efficiencies as for the native state for all three FRET pairs, confirming the correct refolding of the triple-labeled DM-MBP (Figure 6B-D). When DM-MBP was denatured in 3 M GuHCl, the FRET efficiency BG and BR was found to be centered at a value of ~0, while the FRET efficiency GR was at ~0.35 (Figure 6 B-D). To confirm the specific labeling of the dyes and to validate the results with the dyes used for the three-color FRET experiments, we additionally measured double-cysteine mutants probing NTD, CTD and N-C interface separately with two-color smFRET measurements (Supplementary Figure 10A-C) using the respective dye-pairs.

**Three-color FRET on DM-MBP revealed the sequential domain-wise folding of intermediate state towards the native state**

As before, we performed equilibrium unfolding and refolding measurements using triple-labeled DM-MBP. Consistent with the results of the two-color smFRET experiments (Figures 2-3), a similar unfolding trend was observed (Figure 6E). Next, we probed the equilibrium refolding, obtaining a similar trend as compared to the two-color FRET experiments (Figure 2-3 and Figure 6F). For the histogram of the FRET efficiency between the blue and green dye, an intermediate population with $E_{BG} = 0.7$ was evident at concentrations of 0.1-0.5 M GuHCl, in agreement with the intermediate state seen for the NTD (Figure 2) and the two-color FRET controls using the blue-green dye pair (Supplementary Figure 10D-E, left panels). Similarly, an intermediate state was present for the CTD and N-C interface with $E_{GR}$ ~0.8 and $E_{BG}$~0.7, respectively, which was confirmed by two-color FRET controls with the green-red and blue-red dye pairs, respectively (Supplementary Figure 10D-E, middle and right panels). In summary, the three-color FRET measurements confirmed the global intermediate state spanning both the domains and inter-domain interface.

A question that could not be addressed by the two-color FRET experiments is whether transition from the folding intermediate to the native state occurs independently for the NTD and CTD, or whether the two domains fold cooperatively in the same protein molecule. Dynamic PDA on two-color smFRET gave an evidence for the order of folding but three-color smFRET can resolve these transitions in the same molecule. This information is made available from the three-color smFRET experiment by selecting molecules in the intermediate state for one domain and examining the conformation of the other domain (Figure 7A). We used a concentration of 0.5 M GuHCl where the transition occurs from intermediate to the native state (Figure 6F). Molecules were classified according to their conformation detected in the NTD with molecules in the intermediate state were selected using an upper threshold of $E_{BG}$ ~0.7 and in the native state using a lower threshold of $E_{BG}$ ~0.9 (Figure 6F and Supplementary Figure 10E). The conformation of the CTD and N-C interface were probed using the FRET efficiencies $E_{GR}$ and $E_{BR}$. Molecules found in the intermediate state of the NTD also exhibited the intermediate state for the CTD and N-C interface. On the other hand, the majority of molecules that showed a folded NTD had also reached the native state N-C interface but interestingly CTD still has not folded to the native state (Figure 7A and 7B). Thus, the correlated information available with three-color FRET suggests that the intermediate state transitions to the native state in a sequential fashion where NTD and N-C interface folds first and CTD folding follows later.

**Molecular Dynamic simulations of MBP and DM-MBP unfolding**

To obtain structural insights into the folding of DM-MBP, we performed all-atom molecular dynamics (MD) simulations. It is currently not possible to simulate the folding trajectory starting from the denatured state on a realistic time scale. However, unfolding can be induced by increasing the temperature, giving insights into the folding pathways of proteins (Lazaridis & Karplus, 1997). Assuming microscopic reversibility of the folding and unfolding processes, the temperature-induced unfolding would corresponds to the reversed folding pathway (Dinner & Karplus, 1999),(Toofanny & Daggett, 2012). In the case of WT-MBP, which folds within a minute, it was shown that the NTD folds before the CTD (Walters, Mayne, Hinshaw, Sosnick, & Englander, 2013). Here, for DM-MBP, we found the same folding order but delayed on the timescale of ~20-30 min (Figure 1C) (Tang et al., 2006). To investigate the origin of these differences, we simulated the temperature-induced unfolding of WT-MBP and DM-MBP with all atom MD simulations starting from the native conformation (PDB: 1OMP). Simulations were carried out for 2 µs time at 450 K after equilibrating the protein for 2 µs at 400 K. Indeed, the unfolding trajectory for WT-MBP shows that the majority of the secondary structures in the CTD unfolds first, while, for the NTD, parts of the secondary structure are preserved until the end of the simulation (Figure 7C). These regions include two alpha-helices (15-25 amino acid residues and 280-286 amino acid residues) and two-anti-parallel beta sheets (Figure 7F). This suggests that, when WT-MBP folds, these two helices and two- beta sheets are the first structural elements in the NTD and the rest of the CTD folds later. This folding nucleus is disrupted by the double mutations introduced by the DM-MBP. The root-mean-square-deviation (RMSD) of the backbone atoms for WT-MBP unfolding shows step-wise changes, implying step-wise unfolding events (Figure 7E, left panel). We ran two repeats for WT-MBP for unfolding simulations directly at 450 K for 2 µs without the initial 2 µs equilibration run at 400 K (Supplementary Figure 11A-B), which showed the same trend of unfolding events for the NTD and CTD. These results are in agreement with the folding order measured in previous studies on WT-MBP (Walters et al., 2013), (Ye et al., 2018). Strikingly, for DM-MBP, one global unfolding event was observed, and no major secondary structure was preserved at the end of the simulation (Figure 7D). In addition, the RMSD for DM-MBP unfolding also indicates one major unfolding event within 500 ns of the temperature shift from 400 K to 450 K (Figure 7E, right panel). Repeating the simulations run directly at 450 K also lead to complete unfolding in a single step within 300 ns (Supplementary Figure 11C-D). This clearly suggests a decreased stability as the loss of native contacts was higher and faster in DM-MBP.

Thus, results from MD simulations on WT-MBP correctly captured the folding order showing that the NTD folds first followed by the CTD. This shows that the order of folding is conserved from WT- to the double mutant version but with a significant delay in the latter case. However, while the NTD folds faster in WT-MBP, the CTD fluctuates quickly between the folded and unfolded conformations in DM-MBP and can lock into the folded state once the NTD domain finishes folding.

**GroEL cavity and GroEL\ES confinement modulates the dynamics in the folding landscape of DM-MBP**

Molecular chaperones are key players in the process of protein folding. They help nascent chains as well as denatured and aggregated proteins to fold in their final native structure (Balchin, Hayer-Hartl, & Hartl, 2016). The bacterial chaperonin, GroEL\ES, is one of the most widely studied system to understand the mechanism of chaperone assisted protein folding. Previously, it has been shown that the GroEL\ES cavity has effect in accelerating DM-MBP spontaneous folding by 8-13 fold (Tang et al., 2006) (Sharma et al., 2008). In the current study, we have elucidated the underlying fast dynamics of DM-MBP, where the protein fluctuates between near a native state and the unfolded configuration. We asked the question how the GroEL\ES cavity influences the folding dynamics in DM-MBP. Here, we analyzed the initial 10 minutes of DM-MBP folding of the NTD spontaneously as well as in the cavity of GroE\ES. During spontaneous folding, a broad FRET efficiency distribution was observed that exhibits dynamic fluctuations between near native and unfolded state (Figure 8A). When DM-MBP is bound to GroEL in the absence of GroES and ATP, the majority of the DM-MBP is still dynamic between the same of similar states but with an equilibrium shifted towards the unfolded state. In addition, a new static conformation is observed where DM-MBP is further stretched as observed for unfolded DM-MBP conformation (Figure 8B), as observed previously by Sharma et al. (Sharma et al., 2008). Strikingly, a dynamic intermediate is still observed within the cavity of GroEL\ES with similar endpoints observed in the fluorescence lifetime distribution, but the equilibrium is significantly shifted to the compact, high FRET native-like conformation. This suggests that GroEL\ES induces native like conformation in DM-MBP within 10 minutes as previously found, but this study reveals that DM-MBP is still dynamic in the GroEL\ES cavity. However, the chaperonin cavity limits the conformational landscape of the protein, decreasing the entropy of finding the correct native-like structure and thus increasing the folding rate (Figure 8C), as also seen in Sharma et al.

**DISCUSSION**

Spontaneous folding of MBP has been studied for over three decades to investigate different facets of the protein folding process (Spurlino et al., 1991),(Chun et al., 1993),(Raffy, Sassoon, Hofnung, & Betton, 1998),(Bertz & Rief, 2008),(Walters et al., 2013),(Selmke et al., 2018).

MBP and its slow-folding-mutants, SM-MBP (Y283D) and DM-MBP (V8G and Y283D) have been extensively used as model substrates for chaperone assisted folding studies (H Sparrer, Rutkat, & Buchner, 1997),(J. D. Wang, Michelitsch, & Weissman, 1998),(Tang et al., 2006). Importantly, the unfolded states and folding intermediates of WT-MBP and MBP folding mutants show similar features under *in vivo* and *in vitro* conditions, as they both serve as substrate models for different chaperone systems (Mapa, Tiwari, Kumar, Jayaraj, & Maiti, 2012a), (Bandyopadhyay et al., 2012) (Huang, Rossi, Saio, & Kalodimos, 2016). Some of the *E. coli* chaperone systems studied with WT-MBP and DM-MBP are SecB (part of translocation machinery)(Bechtluft et al., 2007), trigger factor (ribosome-associated chaperone)(Mashaghi et al., 2013), DnaK/DnaJ/GrpE (cytoplasmic Hsp40/Hsp70 chaperone system)(Mashaghi et al., 2016) and GroEL/ES chaperonins (Hsp60/Hsp10 chaperonins)(Sharma et al., 2008),(Chakraborty et al., 2010),(Gupta, Haldar, Miličić, Hartl, & Hayer-Hartl, 2014),(Ye et al., 2018).

In three-color smFRET, three spectrally distinct fluorescent dyes are attached to the specific sites on a protein or on multiple interacting partners to obtain three FRET efficiencies at a time. In general, three separately measured FRET efficiencies can also serve the purpose to get the same information, but three-color smFRET has an advantage to have simultaneous information of three distance vectors (Clamme & Deniz, 2005), (Gambin & Deniz, 2010). Importantly, simultaneous information gives access to observe the degree of co-ordination among them. This information is useful, especially for biological systems (Ratzke et al., 2014), (Barth, Voith Von Voithenberg, & Lamb, 2019), (Voith von Voithenberg et al., 2021). Thus, three-color smFRET has opened the possibility to study folding dynamics in multiple domains of large proteins.

In this report, we have investigated the DM-MBP domain-wise folding by considering each domain and the domain interface separately with both by two-color and three-color smFRET. Two-color smFRET provided the information about the conformational heterogeneity with clearly separated unfolded, intermediate and native states in all the three co-ordinates during DM-MBP refolding. We showed that the previously identified intermediate state is indeed not a kinetically-trapped state, but originates from rapid folding-unfolding dynamics on the micro-millisecond time-scale. Additionally, strong evidence provided by dynamic PDA confirms the belief that NTD folds first and CTD folds later. Though, the evidence for co-ordination between all the three vectors was not possible with two-color smFRET analysis in the same molecule; uniquely, three-color smFRET filled this gap by providing the unpreceded details about the correlative motions in domains in DM-MBP folding.

Since the inception of first studies on SM-MBP and DM-MBP as a substrate in the context of GroEL/ES chaperonins, there has been significant interest to understand the effect of mutations on spontaneous folding of MBP (Helmut Sparrer, Lilie, & Buchner, 1996), (J. D. Wang et al., 1998). WT-MBP folds within a minute ($t_{1/2}$ ~25 s) while SM-MBP ($t_{1/2}$ ~3 min) and DM-MBP ($t_{1/2}$ ~30 min) mutants folding is delayed (Tang et al., 2006). Subsequent studies revealed with advent of single molecule FRET methods, that DM-MBP stays in a long-lived intermediate stat refolding is delayed after diluting out the denaturant, which has been attributed to the population of long-lived intermediate states. Characteristic properties shown by the hypothesized long-lived intermediate state, e.g. the exposition of hydrophobic patches, presumably explained its strong interaction with DnaK/J/E and GroEL/ES chaperones (Sharma et al., 2008) (Mapa et al., 2012a). The long-lived intermediate has been thought to be the origin of the hysteresis in the unfolding-refolding curve (Figure 1C) (Chakraborty et al., 2010). Often, *In vitro* folding studies on complex systems can result into aggregation showing hysteresis effect (Moon, Kwon, & Fleming, 2011). We ruled out this possibility in case of hysteresis present in DM-MBP refolding by confirming absence of any reversible aggregates with FCCS (Figure 1D). In this study we have characterized the hysteresis in DM-MBP, with both two and three-color smFRET. Certainly, the cause of the of hysteresis can be attributed to the properties of the conformations displayed by DM-MBP refolding molecules between 0.4-~1 M GuHCl concentrations, as seen previously as well (Figure 2-3, Figure 6) (Chakraborty et al., 2010). It has been speculated that the double mutations present in DM-MBP, delay the formation of nucleation core in NTD, a rate limiting step in overall folding of a protein (Chun et al., 1993). NTD core encompasses the residues which are far apart in sequence, but juxta positioned in the native state. Disruption of this core as a result of two point-mutations with less hydrophobic side chains (V8G, Y283D), led to increased flexibility of the segment containing two helices and two beta sheets (Figure 7C-F) (Ye et al., 2018). Constraining this flexible region by introducing disulfide bridge tethering the distant segments found to remove the hysteresis by decreasing the entropic barrier to form native-like contacts. Furthermore, usage of a small molecule known to reduce structure flexibility in protein loops enhances the folding of DM-MBP (Figure 5E) (Bandyopadhyay et al., 2012). This indicates that DM-MBP encounters an entropic barrier leading to a local minimum in the folding funnel, which the protein has to overcome to attain the native state (Figure 4D) (Figure 5D). The resulted barrier can be overcome e.g. with the help of a chaperonin cage that decreases the entropic barrier by confinement (Gupta et al., 2014), (Tang et al., 2006), (Chakraborty et al., 2010).

A hydrogen exchange-mass spectroscopy (HX-MS) study on WT-MBP revealed that the NTD folds first ($t_{1/2}$ ~1 s) while the CTD folds later ($t_{1/2}$ ~40 s) (Walters et al., 2013). Subsequent HX-MS study with one of the mutant V8G in MBP demonstrated the delay in NTD folding ($t_{1/2}$ ~20

s) and no effect on CTD folding ($t_{1/2}$ ~40 s) (Ye et al., 2018), implying that the V8G mutation doesn't impart the dependency of the CTD on the NTD folding. In the current study, we confirmed the dependency between the folding of the NTD and CTD in the presence of the additional Y283D mutation, limiting the overall folding of DM-MBP ($t_{1/2}$ ~30 min), as shown with the analysis of the folding kinetics by dynamic PDA (Figure 5A-C) (Table 5). Remarkably, three-color smFRET measurements allowed us to follow the folding of both the domains and domain interface simultaneously. This provided strong evidence for implicated intermediate state, spanning both the domains and inter-domain interface, dependent folding of the NTD and CTD in the same molecule (Figure 7A-B). The correlated information obtained from three-color smFRET provided direct evidence for the order of the domain folding, showing that the folding of the NTD is necessary for the CTD to fold. This demonstrates how certain mutation affects the domain-wise folding as well as the extend of domain dependency on overall folding as a result of sequence discontinuity in domains imposing a significant entropic barrier of 9.7 kJ/mol (Figure 5D). Furthermore, it recapitulates the evolution of cooperative domain folding having discontinuity in amino acid sequence (Vogel, Bashton, Kerrison, Chothia, & Teichmann, 2004) (Han et al., 2007) (Inanami, Terada, & Sasai, 2014) (Y. Wang, Chu, Suo, Wang, & Wang, 2012) (Arviv & Levy, 2012), having implications in preventing aggregation between the domains and subsequent misfolding stress (Borgia et al., 2011), (Mashaghi et al., 2013) (Cerminara, Schöne, Ritter, Gabba, & Fitter, 2020). Remarkably, to achieve the task of complex multi-domain folding, cells might have to depend occasionally on ribosomes and chaperones to attain native structures in physiological time-scales (Holtkamp et al., 2015) (Agashe et al., 2004) (Liu, Maciuba, & Kaiser, 2019) (Imamoglu, Balchin, Hayer-Hartl, & Hartl, 2020). This also supports the notion of co-evolved chaperone-substrate networks. We found that DM-MBP bound to GroEL alone and in the cavity of GroEL\ES is still mostly dynamic, but the environment reshapes the underlying dynamics and explored conformational space. The hydrophobic lining of GroEL captures the unfolded substrate and shifts the equilibrium conformation slightly towards the unfolded state whereas the hydrophilic cavity of GroEL\ES limits the conformational space explored by the protein and allows it to move downhill in the folding funnel. This shows the capability of chaperones to tune the folding landscape of a substrate protein and is one mechanism to overcome entropic barriers in multi-domain proteins.

In summary, the presented study illustrates how the folding of complex multi-domain proteins can be studied using two- and three-color smFRET. Using this workflow, the generality of the folding pathways can be tested using well studied systems like adenylate kinase and phosphoglycerate kinase with complex domain topologies (Osváth, Köhler, Závodszky, & Fidy, 2005) (Li, Terakawa, Wang, & Takada, 2012). Certainly, this is possible with improved spectral choices for multi-color smFRET and robust labeling approaches to achieve specific labeling of more than two fluorophores. Eventually, it would be interesting to see how known interacting chaperone systems of DM-MBP like DnaK/J/E and GroEL/ES modulates the domain-wise folding of DM-MBP with the feasibility to perform three-color smFRET measurements. This can provide the difference in general folding mechanisms adopted by these two folding systems which might be applicable to speculate about folding pathways of their specific substrates.

## MATERIALS AND METHODS

**MBP constructs, protein expression and purification**

Plasmid encoding DM-MBP (V8G, Y283D) was a generous gift from F. Ulrich Hartl and Manajit Hayer-Hartl (MPI of Biochemistry, Martinsried, Germany). The DM-MBP plasmid backbone is pCH-series based, (Chang, Kaiser, Hartl, & Barral, 2005), enabling IPTG inducible expression. Single cysteine (A52C, P298C), double cysteine (A52C-P298C, K175C-P298C, A52C-K175C) DM-MBP mutants were generated using site-directed mutagenesis (Thermo Scientific-Phusion Site-Directed Mutagenesis Kit). The A52TAG-K175C-P298C mutant was generated by mutating the codon for alanine at position 52 to the amber stop codon (TAG) to incorporate the unnatural amino acid (UAA) N-Propargyl-L-Lysine (PrK).

All single and double cysteine DM-MBP mutant proteins were expressed in *E. coli* BL21-AI (L-(+)-arabinose controlled T7 RNA polymerase expressing strain) at 30° C for 4 hr, 200 RPM with the addition of both 0.2% L-(+)-arabinose and 0.5 mM IPTG.

UAA incorporation requires the co-expression of the used orthogonal translation system of tRNA$^{Pyl}$ and PylRS$^{WT}$ with the expression of the given protein. For this purpose, we used pEvol-tRNA$^{Pyl}$PylRS$^{WT}$ (Chatterjee, Xiao, & Schultz, 2012). *E. coli* BL21-AI cells harboring both pCH-A52TAG-K175C-P298C-DM-MBP and pEvol-tRNA$^{Pyl}$PylRS$^{WT}$ plasmids were grown for 1 hr at 30° C before the addition of 1mM of PrK (SiChem GmbH) in the media. PrK was prepared in 0.1 M NaOH solution. Later, DM-MBP-A52PrK-K175C-P298C expression was achieved similar to the other mutants with L- (+)-arabinose and IPTG.

All DM-MBP proteins were purified with an amylose column (New England Biolabs) as described previously (Sharma et al., 2008). Proteins were quantified spectrophotometrically at 280 nm.

**Tryptophan fluorescence measurements**

All the tryptophan fluorescence measurements were performed at 20 °C on FLS1000 Photoluminescence Spectrometer (Edinburgh Instruments). To investigate the kinetics of MBP refolding, intrinsic tryptophan fluorescence measurements were performed on refolding MBP. Fluorescence intensity was measured for 2 s at intervals of 60 s. 3 µM MBP was denatured with 3 M GuHCl in buffer A (20 mM Tris, pH 7.5, 20 mM KCl) and allowed to refold after 75-fold dilution in buffer A. Intrinsic tryptophan fluorescence was excited at 290 nm with a slit width of 2 nm and detected at 345 nm with a 5 nm slit width. Photobleaching was avoided by adjusting the slit widths, measurement time intervals and acquisition times.

For both unfolding and refolding curves, steady-state tryptophan fluorescence was measured after 20 hr. For the unfolding curve, ~40 nM native MBP was incubated at 22 °C in buffer A containing 0.2 M, 0.4 M, 0.6 M, 0.8 M, 1 M, 1.2 M, 1.4 M, 1.6 M, 1.8 M and 2 M GuHCl. For the refolding curve, 2 µM MBP was denatured in 3 M GuHCl/10 mM DTT at 50° C for 1 h in buffer A and incubated at 22 °C after 50-fold dilution in 0 M, 0.2 M, 0.4 M, 0.6 M, 0.8 M, 1 M, 1.2 M, 1.4 M, 1.6 M, 1.8 M and 2 M GuHCl prepared in buffer A.

**The Boltzmann function for unfolding and refolding titrations**

The unfolding and refolding titrations from the tryptophan measurements were fit using a Boltzmann function:

$$y = \frac{A_1 - A_2}{1 + e^{(I-I_0)/\Delta I}} + A_2$$

where $A_1$ is the final and $A_2$ is the initial data point respectively, $I$ is the center of the transition and $\Delta I$ is the increment between data points.

For equilibrium unfolding and refolding experiments with smFRET, an intermediate population was observed. In these cases, the titrations were fit using a double Boltzmann function given by:

$$y = y0 + \frac{f_1}{1 + e^{(I-I1_0)/\Delta I_1}} + \frac{(1-f_1)}{1 + e^{(I-I2_0)/\Delta I_2}}$$

where $f_1$ is the amplitude of the first transition, $I1_0$ and $I2_0$ are the centers of the transitions and $\Delta I_1$ and $\Delta I_2$ are the respective increments between data points.

**Fluorophore labelling of MBP**

*Cysteine-maleimide labelling with one and two fluorophores*

All cysteine-maleimide couplings were performed according to the manufacturer's instructions (Atto-Tec) with a few modifications. The single-cysteine mutant, A52C was labelled either with Atto532- (Atto-Tec) or Alexa647-maleimide (Invitrogen). All double-cysteine mutants, A52C-P298C, K175C-P298C and A52C-K175C were stochastically labelled with Atto532- and Alexa647-maleimides. To analyze two-color controls of three-color FRET analysis, the double-cysteine mutant A52C-P298C was stochastically labelled with Atto488- and Atto565-maleimides, K175C-P298C with Atto565- and Alexa647-maleimides and, A52C-K175C with Atto488- and Alexa647-maleimides.

Briefly, the sulfhydryl groups of ~50 µM DM-MBP cysteine mutants were reduced with 10 mM DTT in phosphate buffered saline solution (PBS) at room temperature (RT) for 20 min. Excess DTT was removed by washing the protein in a 10 kDa cut-off Amicon centrifugal filter (Merck-Millipore) with de-oxygenated PBS containing 50 µM tris(2-carboxyethyl) phosphine (TCEP). Approximately a 3-fold molar excess of the maleimide-fluorophore conjugate was added to the washed protein solution and the reaction was carried out at RT for 3 hr in the dark. For double labelling, the fluorophores were labeled stochastically by adding an equimolar mixture of both maleimide-fluorophore conjugates simultaneously in the reaction. Unreacted maleimide-fluorophores were washed out with buffer A containing 1 mM DTT using a centrifugal filter. Successful labeled was verified using FCS. FCCS was used to confirm the double-labelling. The degree of labelling was quantified spectrophotometrically. Labeled of DM-MBP with the above mentioned cysteine-maleimide chemistries did not affect the folding rates (Table 1).

*Specific labelling with three fluorophores*

DM-MBP-A52PrK-K175C-P298C protein was specifically labelled with three fluorophores for the three-color smFRET experiments. In the first step, A52PrK with an alkyne group was specifically conjugated to the azide moiety of Atto488-azide (Atto-tec) via copper-catalyzed alkyne-azide cycloaddition, one type of click chemistry reaction (Chatterjee et al., 2012). ~120 µM of DM-MBP-A52PrK-K175C-P298C protein was allowed to react with 3-fold molar excess of Atto488-azide, in the presence of 200 µM $CuSO_4$, 50 µM TCEP, 200 µM TBTA and freshly prepared 200 µM sodium ascorbate in PBS at RT for ~3 hr, in the dark, under mild shaking conditions (Tyagi & Lemke, 2013). Unreacted dye was removed by washing with PBS using centrifugal filters.

K175C and P298C were labelled with Alexa647 and Atto565, respectively, using cysteine-maleimide chemistry. It has been shown that maltose binding in the inter-domain cleft of MBP buries some residues at the domain-interface including 298 position (Sharma et al., 2008) (Mapa, Tiwari, Kumar, Jayaraj, & Maiti, 2012b). Consistent with this observation, we labelled the Alexa647-maleimide specifically to the cysteine at 175 position by labeling in the presence of maltose, which blocks the competing cysteine of 298 position (Sharma et al., 2008), (Mapa et al., 2012b) (Jäger, Michalet, & Weiss, 2005). For the second step of labelling, we took ~70 µM of 52PrK-atto488 labelled DM-MBP-A52PrK-K175C-P298C, reduced with 10 mM DTT addition and washed with de-oxygenated PBS containing 500 mM Maltose. The cysteine-maleimide reaction was performed by the addition of 2-fold molar excess of Alexa647-maleimide and allowed to react for 1 hr at RT to minimize the possibility of mis-labelling the maltose blocked cysteine at position 298 (Supplementary Figure 9). As a third and final step, ~30 µM of 52PrK-Atto488-175C-Alexa647 labelled DM-MBP-A52PrK-K175C-P298C protein was washed with PBS to remove the excess of unlabeled Alexa647 dye and maltose. The washed protein was then labelled with 3-fold molar excess of Atto565-maleimide to the only available cysteine at position 298. Coupling of each dye after each labelling step was monitored by measuring the absorption of the respective fluorophores at their respective wavelengths of maximum absorption and of the protein at 280 nm. The overall degree of labeling for all three labels and was estimated to be ~15% by absorption spectroscopy. However, on the lower limit ~2 % triple-labeled molecules were finally used to analyze the FRET histograms after filtering for photobleaching and blinking events (Supplementary Figure 8B-C). Covalent attachment of the three-fluorophores did not have any significant influence on the folding (Figure 6B-D, Supplementary Figure 8F).

**Single molecule measurements and data analysis**

*Two-color and three-color FRET measurements*

All the FRET measurements were performed on custom -build confocal set-ups as described below. 50-100 pM of double- or triple-labelled DM-MBP proteins were measured to minimize the possibility of having more than one molecule in the confocal volume at a time. Before starting a smFRET measurement on the glass slide, the surface was passivated with 1 mg/ml BSA. SmFRET measurements of native MBP proteins were performed after serially diluting the labelled proteins in buffer A (20 mM Tris, pH 7.5, 20 mM KCl). To measure the FRET efficiency in the completely unfolded state, first ~500 nM labelled protein was denatured in buffer A containing 3 M GuHCl, 10 mM DTT at 50° C for 1 h and later was measured after serially diluting the protein concentration to picomolar 50-100 pM in 2 or 3 M GuHCl prepared in buffer A. SmFRET experiments performed in 6 M GuHCl showed no additional changes in the conformation of the denatured protein.

For all unfolding and refolding FRET measurements, 0.001% tween-20 was added to the buffer to prevent unfolded and refolded molecules from sticking to the surface. To measure the FRET efficiency under unfolding conditions, labelled native protein was diluted to 50-100 pM in buffer A containing either 0 M, 0.1 M, 0.2 M, 0.3 M, 0.5 M, 0.9 M or 1 M GuHCl. Refolding FRET measurements were performed by first denaturing 500-1000 nM protein in 3 M GuHCl/10 mM DTT at 50°C for 1 hr followed by serial dilutions in 3 M GuHCl and a final 50-fold dilution in buffer A/GuHCl to obtain 50-100 pM labeled protein in buffer with the desired GuHCl concentration.

*Two-color setup*

Fluorescence correlation and cross-correlation spectroscopy and for two-color smFRET measurements on DM-MBP proteins labelled with Atto532 and Alexa647 were performed on a home-built confocal microscope capable of multi-parameter fluorescence detection (MFD) combined with pulsed interleaved excitation (PIE, MFD-PIE). PIE was implemented using a 532-nm green laser (Toptica; PicoTA 530) to excite the donor and a 640-nm red laser (PicoQuant; LDH-D-C-640) for directly exciting the acceptor molecules. The lasers were synchronized to a repetition rate of ~26.7 MHz with a delay of ~18 ns between each pulse. Laser excitation powers were set to ~100 µW before the objective for both the lasers. A 60x water immersion objective (Nikon; Plan Apo IR 60x1.27 Water Immersion) was used to collect the emitted fluorescence and focused onto a 75µm diameter pinhole for confocal detection.

To implement MFD, the fluorescence signal collected after the pinhole was separated into parallel and perpendicular polarized light by a polarizing beam-splitter (Thorlabs; PBS3) and then spectrally separated for green and red fluorescence by a dichroic mirror (AHF Analysetechnik; Dual Line z532/635,). Finally, before fluorescence detection using four single-photon-counting (SPC) avalanche photodiodes (APD) (Perkin-Elmer), both the green and red fluorescence spectra were cleaned with emission filters (green: Semrock, Bright line 582/75; red: Chroma, HQ700/75 M). SPC cards (Becker and Hickl; SPC 154) were synchronized with the laser drivers to record the arrival times of the photons.

*Fluorescence Correlation and Cross-correlation Spectroscopy*

Fluorescence correlation spectroscopy (FCS) and fluorescence cross-correlation spectroscopy (FCCS) experiments were performed with PIE capabilities to investigate aggregation of DM-MBP during refolding (Müller et al., 2005). For these experiments, a single-cysteine mutant of DM-MBP (A52C) labeled with either Atto532 maleimide or Alexa647 maleimide dye was used and measured with the two-color MFD-PIE setup as described above. FCCS detects the coincidence of fluctuations in both the green and red detection channels and hence sensitively detector whether oligomers have formed during. For these experiments, 500 nM Atto532-DM-MBP and 500 nM Alexa647-DM-MBP were denatured in 3 M GuHCl at 50°C for 1 hr. The samples were then mixed at RT and refolding was initiated by diluting the solution to a final labeled-protein concentration of 2 nM (1 nM of Atto532-DM-MBP and 1 nM of Alexa647-DM-MBP) in buffer A. FCCS experiments were performed during the initial 2 minutes after dillution and after 30 to monitor the time dependent oligomer formation over the entire refolding process. A 40 base-pair dsDNA labeled with both Atto532 and Atto647 on separate strands was used as a positive control. Free dyes were measured as a negative control (Figure 1D).

A 3D Gaussian confocal volume was assumed yielding the following analytical equation for the auto- and cross-correlation functions and was used to analyze the FCS and FCCS data:

$$G(\tau) = \frac{\gamma}{N} \frac{1}{\left(1 + \frac{\tau}{\tau_D}\right)} \frac{1}{\sqrt{\left(1 + \frac{\tau}{\tau_D}\frac{1}{\rho^2}\right)}} + y_0$$

where, $\gamma = 2^{-3/2}$ is the geometric factor, to correct for confocal shape, *N* is the average number of diffusing molecules in the probe volume with a diffusion time of $\tau_D = \frac{w_0^2}{4D}$, where $D$ is the diffusion coefficient, $\rho$ is the structure parameter defined as $w_0/z_0$, where $w_0$ and $z_0$ are the axial and radial dimensions from the center of the point-spread-function to the position where the intensity has decayed to $\frac{1}{e^2}$ and $y_0$ is a baseline to compensate for a potential offset in the correlation functions. For the FCCS analysis, the amplitudes of the green and red autocorrelation functions can be used to determine the total number of diffusing particles containing a green label and a red label:

$$N_{GT} = N_G + N_{GR} \text{ and } N_{RT} = N_R + N_{GR}$$

where $N_G$ and $N_R$ represent the number of diffusing particles containing only a green and a red label respectively. The number of double-labeled molecules ($N_{GR}$) was determined from the amplitude of the cross-correlation function, which is given by:

$$G_{CC}(0) = \frac{\gamma \times N_{GR}}{N_{GT} \times N_{RT}}$$

*Two-color MFD-PIE analysis*

For two-color smFRET measurements, ~50-100 pM double labeled MBP sample was measured. PIE-MFD data analysis was performed to calculate correct two-color FRET efficiency, burst-wise lifetimes and anisotropy for single molecule events (Supplementary Figure 1) as described previously (Zander et al., 1996) (Kudryavtsev et al., 2012). Single molecule bursts in two-color FRET measurements were distinguished from the background with a use of photon burst search algorithm by applying a threshold of at least 5 photons for sliding time window of 500 µs and a total of 50 photons per burst (Nir et al., 2006). The burst-wise fluorescence lifetime was estimated from the florescence decay by reconvolution with the instrument response function. The fluorescence anisotropy was fitted with respect to the fluorescence lifetime using Perrin equation as shown previously (Schaffer et al., 1999):

$$r = \frac{r_0}{1 + \frac{\tau}{\rho}},$$

where, $r$ is steady state anisotropy, $r_0$ is the fundamental anisotropy, $\rho$ is rotational correlation time and $\tau$ is a fluorescence lifetime. Molecules labelled with both donor and acceptor dyes showing a stoichiometry of ~0.5 were selected for further analysis. An ALEX-2CDE filter with an upper value of 12 was used to filter out photobleaching and blinking events (Tomov et al., 2012). After correcting for background, crosstalk of green fluorescence in red channel ($\alpha$), direct excitation of acceptor by donor excitation laser ($\delta$), and differences in detection efficiencies and quantum yields of the dyes ($\gamma$) were estimated. The corrected labelling stoichiometry (S) and FRET efficiency for all the bursts was calculated as:

$$\left[ S = \frac{\gamma F_{GG} + F_{GR} - \alpha F_{GG} - \delta F_{RR}}{\gamma F_{GG} + F_{GR} - \alpha F_{GG} - \delta F_{RR} + F_{RR}} \right]$$

$$\left[ E = \frac{F_{GR} - \alpha F_{GG} - \delta F_{RR}}{\gamma F_{GG} + F_{GR} - \alpha F_{GG} - \delta F_{RR}} \right]$$

where, $F_{GG}$ and $F_{GR}$ are the fluorescence signal in donor and acceptor channel after donor excitation respectively and $F_{RR}$ is the fluorescence signal in red channel after acceptor excitation. For the determined correction factors, see Table 1.

Ideally, when the conformational of the protein is static while the molecule transits the laser spot, the FRET efficiency (E) is related to the fluorescence lifetime of the donor in presence of an acceptor as:

$$\left[E = 1 - \frac{\tau_{D(A)}}{\tau_{D(0)}}\right]$$

where $\tau_{D(A)}$ is the fluorescence lifetime of donor in presence of an acceptor and $\tau_{D(0)}$ is the fluorescence lifetime of donor in absence of an acceptor. This relationship changes slightly when the inter-dye separation becomes comparable to the relative linker lengths, where linker flexibility dominates. An accurate relationship can be derived for a specific set of fluorophores as has been described previously (Gansen et al., 2009). For the dye pair Atto532-Alexa647, this relationship is given by the following third-order polynomial:

$$\left[E = 1 - \frac{-0.0178 + 0.6226 \langle\tau_{D(A)}\rangle + 0.2188 \langle\tau_{D(A)}\rangle^2 + 0.0312 \langle\tau_{D(A)}\rangle^3}{\langle\tau_{D(0)}\rangle}\right]$$

When dynamics are present between two states with their respective donor fluorescence lifetimes $\tau_1$ and $\tau_2$, the relationship of intensity averaged FRET efficiency (E) to the donor lifetime changes to (Kalinin, Valeri, et al., 2010):

$$\left[E = 1 - \frac{\tau_1 \cdot \tau_2}{\tau_{D(0)}[\tau_1 + \tau_2 - \langle\tau\rangle]}\right]$$

where, the average donor fluorescence lifetime $\langle\tau\rangle$ is calculated from the total florescence signal over a burst.

*Fluorescence lifetime analysis*

To determine the FRET efficiencies of the different FRET states undergo dynamic transitions, a fluorescence lifetime analysis was performed. The photons from all bursts were sum together and fit to a biexponential function convoluted with the instrument response function and a scattering component was included as an additional species in the fit. From the fluorescence lifetime, the FRET efficiency and distances were calculated using:

$$\left(E = 1 - \frac{\tau_{D(A)}}{\tau_{D(0)}}\right) \text{ and } \left(E = \frac{1}{1 + \left(\frac{R}{R_0}\right)^6}\right)$$

where $\tau_{D(A)}$ is the fluorescence lifetime of donor in presence of an acceptor, $\tau_{D(0)}$ is the fluorescence lifetime of donor in absence of an acceptor (~3.6 ns for Atto532), R is the distance between the dyes and $R_0$ is the Förster radius (62 Å for Atto532-Alexa647 dye pair). The quality of the fit was evaluated using the reduced $\chi^2$ value, $\chi^2_{red}$.

*Three-color setup*

Both triply-labelled DM-MBP with Atto488-Atto565-Alexa647 and doubly-labelled DM-MBP with Atto488-Atto565, Atto565-Alexa647 and Atto488-Alexa647 dye pairs were measured on a three-color confocal single molecule setup equipped with MFD-PIE as previously described (Barth, Voith Von Voithenberg, & Lamb, 2019). Briefly, PIE experiments were performed with three pulsed lasers having ~20 ns delay between each pulse (PicoQuant, Germany; LDH-D-C-485, LDH-D-TA-560, LDH-D-C-640). The pulse frequency of 16.7 MHz and its synchronization was achieved using a laser driver (PicoQuant, Germany; Sepia II). A 60x water immersion objective with 1.27 N.A. (Nikon, Germany; Plan Apo IR 60x 1.27 WI) was used to focus the lasers into the sample with a power measured before the objective of ~120 µW for blue, ~75 µW for green and ~35 µW for red laser. The emitted fluorescence was collected by the same objective and separated from the laser excitation using a polychroic mirror (AHF Analysentechnik; zt405/488/561/633, Germany) and passed through 50 µm pinhole for defining the confocal volume. Light coming through the pinhole was first separated for its parallel and perpendicular polarization with a polarizing beam splitter (Thorlabs, Germany). Afterwards, the light in each polarization channel was separated into blue, green and red spectral regions using two dichroic mirrors (AHF Analysentechnik; BS560, 640DCXR). The blue, green and red detection channels were spectrally clean using emission filters (AHF Analysentechnik; ET525/50, ET607/36, ET670/30) before the fluorescence was detected on APD's (LaserComponents, 2x COUNT-100B; Perkin Elmer, 4x SPCM-AQR14). The timing of the detected photons was synchronized with lasers pulses using a TCSPC module (PicoQuant; HydraHarp400).

*Three-color MFD-PIE analysis*

In a three-color FRET system, the blue dye acts as a donor (D), the green dye acts as a first acceptor ($A_1$) and the red dye as a second acceptor ($A_2$). Additionally, the green dye can also act as a donor for the red dye (Supplementary Figure 8A). Extending MFD-PIE to three-colors makes it possible to detect photons in green and red channels after blue excitation as well as in the green and red channels after green excitation. This enables one to calculate all three stoichiometries ($S_{BG}$, $S_{BR}$ and $S_{GR}$) and FRET efficiencies ($E_{BG}$, $E_{BR}$ and $E_{GR}$) for blue-green, blue-red and green-red dye-pairs respectively. The latter case of green-red is similar to the typical two-color MFD-PIE scheme.

For three-color FRET measurements, ~50-100 pM triple labelled DM-MBP was measured on a passivated glass surface. An all-photon burst search algorithm was used to detect the single molecule events from the background and required at least 30 photons per sliding window of 500 µs and a total of 100 photons per burst. A typical burst-wise MFD-PIE analysis inclusive of stoichiometry, FRET efficiency, fluorescence lifetime and anisotropy, was extended to three-color MFD-PIE on the selected bursts as explained previously (Supplementary Figure 8) (Barth et al., 2019a).

Briefly, the three stoichiometries ($S_{BG}$, $S_{BR}$ and $S_{GR}$) were calculated as follows:

$$\left[ S_{BG} = \frac{F_{BB} + F_{BG} + F_{BR}}{F_{BB} + F_{BG} + + F_{BR} + F_{GG} + + F_{GR}} \right]$$

$$\left[ S_{BR} = \frac{F_{BB} + F_{BG} + F_{BR}}{F_{BB} + F_{BG} + + F_{BR} + F_{RR}} \right]$$

$$\left[ S_{GR} = \frac{F_{GG} + F_{GR}}{F_{GG} + F_{GR} + F_{RR}} \right]$$

where $F_{XY}$ represents the detected fluorescence signal in the Y channel after exciting the X dye. For the three-color analysis, triple labeled molecules were sorted by applying the ALEX-2CDE-filter for all the three stoichiometries with maximum value of 15 for both blue-green, blue-red and 20 for green-red dye-pairs. Typical values of $S_{BG}$, $S_{BR}$ and $S_{GR}$ for triple-labeled DM-MBP molecules with Ato488, Atto565 and Alexa647 dyes were found to be ~0.2, ~0.15 and ~0.5 respectively (Supplementary Figure 8B-C).

The corrected three FRET efficiencies for selected triple-labeled molecules were derived as detailed in Barth, Voith Von Voithenberg, & Lamb, 2019 and can be written as:

$$\left[ E_{GR} = \frac{F_{GR} - \alpha_{GR} F_{GG} - \delta_{GR} F_{RR}}{\gamma_{GR} F_{GG} + F_{GR} - \alpha_{GR} F_{GG} - \delta_{GR} F_{RR}} \right]$$

$$\left[ E_{BG} = \frac{F_{BG}^{cor.}}{\gamma_{BG} F_{BG} (1 - E_{GR}) + F_{BG}^{cor.}} \right]$$

$$\left[ E_{BR} = \frac{F_{BR}^{cor.} - E_{GR} \left( \gamma_{GR} F_{BG}^{cor.} + F_{BR}^{cor.} \right)}{\gamma_{BR} F_{BB} + F_{BR}^{cor.} - E_{GR} \left( \gamma_{BR} F_{BB} + \gamma_{GR} F_{BG}^{cor.} + F_{BR}^{cor.} \right)} \right]$$

Where, the intermediate correction terms $F_{BG}^{cor.}$ and $F_{BR}^{cor.}$ are defined as:

$$[F_{BG}^{cor.} = F_{BG} - \alpha_{BG} F_{BB} - \delta_{BG} F_{GG}]$$

$$[F_{BR}^{cor.} = F_{BR} - \alpha_{BR} F_{BB} - \delta_{BR} F_{RR} - \alpha_{GR}(F_{BG} - \alpha_{BG} F_{BB}) - \delta_{BG} E_{GR} (1 - E_{GR})^{-1} F_{GG}]$$

The respective crosstalk, direct excitation and detection correction factors are depicted as $\alpha_{XY}$, $\delta_{XY}$ and $\gamma_{XY}$ for signal in channel Y after exciting the dye X.

*Dynamic photon distribution analysis*

The raw photon signal carries important information about the kinetic heterogeneity of the system. For the purpose of computing the interconversion rates between two states in a robust way, first the proximity ratio (*PR*) [($PR = F_{GR}/(F_{GR}+F_{GG})$)] collected during the burst was sliced into 0.5, 1 and 1.5 ms time bins to capture the influence of the kinetics. A global analysis of all the three-time bins was performed to extract the rates after solving it analytically. To take care of the broadening due to photon detection noise, a constant width ($\sigma$) for a static state was assumed to scale with the inter-dye distance (*R*) as $\sigma = 0.07R$ (Kalinin, Sisamakis, Magennis, Felekyan, & Seidel, 2010). The states were defined using the donor fluorescence lifetimes of the double-labeled molecules (See Table 4). Additional states were incorporated to account for impurities and donor only mixing in a sample typically at low proximity ratios. The analysis was applied to extract the rates between the unfolded and folded states in DM-MBP during refolding in various denaturant concentrations and are reported in Table 5.

*Data analysis software*

All the fluorescence correlation, burst analysis, fluorescence lifetime and photon distribution analysis were performed with the open-source PIE Analysis with MATLAB (PAM) software, a custom-written software in MATLAB (The MathWorks) (Schrimpf, Barth, Hendrix, & Lamb, 2018).

**All-atom molecular dynamics simulations**

To obtain atomistic insights into the structural changes of DM-MBP during folding, we performed all-atom molecular dynamics (MD) simulations on both WT- and DM-MBP. For simulations on WT-MBP, the crystal structure with Protein Data Bank (PDB) ID 1OMP (Sharff et al., 1992) was used. For DM-MBP, the amino acids at postion 8 and 283 were exchanged with glycine and aspartate (V8G, Y283D) respectively to create the DM-MBP structure in PyMOL/Chimera (The PyMOL Molecular Graphics system, version 2.0 Schrodinger)/ (Pettersen et al., 2004). The AMBER16 MD package with the ff14SB force field was used for the simulations (Case et al., 2017). TIP3P water model was used to solvate the MBP molecule in a box of octagonal geometry. Care was taken to exclude the vaporization effects at high temperature, as explained earlier (Walser, Mark, & Van Gunstere, 2000) . A distance of 3 nm between the polypeptide chain and the box wall was maintained not to have a boundary effects in simulations. The charge of the system was neutralized by the addition of sodium ions. A small excess of sodium chloride was added, corresponding to a concentration of ~4 mM. Stability of the simulation box was verified at high temperatures at 400/450 K during the initial 500 ps of equilibration using 2 fs steps. MD runs were performed using the NPT ensemble with volume scaling enabled for both the thermostat and a barostat. A Nvidia GTX 1080 Ti GPU was used to carry out the simulations, which typically ran with an average of 50 ns per day. The unfolding trajectory was analyzed using various AmberTools (Roe & Cheatham, 2013). The progress of unfolding was evaluated based on the secondary structure assignment by the DSSP algorithm (Definition of Secondary Structure of Proteins) (Kabsch & Sander, 1983) over the span of the unfolding trajectory at time steps of 5 ns. Furthermore, the fraction of native contacts (Q(X)) was used to evaluate the overall unfolding in the protein (Best, Hummer, & Eaton, 2013).

We performed simulations at both 400 K and 450 K. Simulations at 400 K were run for 2 µs to check the stability of the system at high temperature. Later, simulations were continued for another 2 µs at 450 K. In addition, separate simulations were carried out at 450 K for 1 µs.

**Accessible volume calculations**

To model the FRET distances in the labeled DM-MBP for comparison with the experimentally determined distances for the various labeling positions (PDB ID: 1OMP), we performed geometric accessible volume (AV) calculations with the use of the FRET positioning and screening (FPS) software (Kalinin et al., 2012). The input parameters used for simulating the dye with the AV1 model were: 20 Å (dye linker-length), 4.5 Å (dye width), and 3.5 Å (dye radius).

**Acknowledgements**

We gratefully acknowledge the financial support of the SFB 1035 (German Research Foundation DFG, Sonderforschungsbereich 1035, Projektnummer 201302640, project A11 to D.C.L.) and of the Ludwigs-Maximillians-Universität München through the Center for NanoScience (CeNS) and LMUinnovativ BioImaging Network (BIN). We thank Prof. F. Ulrich Hartl from the Max Planck Institute of Biochemistry, Martinsreid, for kindly providing the GroEL,GroES proteins and for valuable discussions.

**Author contributions**

G.A. purified and labeled proteins, performed all the experiments, analyzed single-molecule data and prepared figures. A.B. implemented and helped in data acquisition, analysis of the three-color smFRET and performed and analyzed molecular dynamics simulations. G.A. wrote the initial version of the manuscript and all authors contributed to the writing of the final version of the manuscript. D.C.L. designed the research.

**Competing interests**

Authors declare that there are no competing interests.

**Table 1:** Refolding rates and half-life ($t_{1/2}$) of refolding for various DM-MBP mutants. All the refolding measurements used for measuring the refolding rates were carried out in GuHCl concentrations below 0.1 M and the data was fitted with a mono-exponential function. Error indicated is a standard deviation from at least three independent measurements.

| | Refolding rate (min$^{-1}$) | $t_{1/2}$ of refolding (min) |
|---|---|---|
| WT-MBP[a] | 2.44±0.53 | 0.29±0.07 |
| DM-MBP[a] | 0.028±0.003 | 24.70±2.65 |
| NTD (52C-298C) Atto532-Alexa647[b] | 0.023±0.0003 | 29.75±0.39 |
| NTD (52C-298C) Atto488-Atto565[b] | 0.030±0.001 | 22.45±1.41 |
| CTD (175C-298C) Atto532-Alexa647[b] | 0.039±0.016 | 19.28±7.88 |
| CTD (175C-298C) Atto565-Alexa647[c] | 0.027±0.011 | 27.41±11.34 |
| N-C interface (52C-175C) Atto532-Alexa647*[c] | 0.031±0.008 | 22.87±6.10 |
| N-C interface (52C-175C) Atto488-Alexa647[b] | 0.035±0.020 | 23.28±13.47 |
| DM-MBP (52PrK-175C-298C) Atto488-Atto565-Alexa647*[c] | 0.031±0.008 | 22.87±6.10 |

[a] Rate was estimated from tryptophan fluorescence measurements.
[b] Rate was estimated from FRET histogram by calculating the increase in the folded fraction relative to the folded fraction for refolded protein.
[c] Rate was estimated by calculating the increase in the fraction of low acceptor lifetime relative to refolded protein (Alexa647 as an acceptor). See Table S1 and Table S2.
*FRET histograms could not be used for these constructs to determine the refolding rate because of low contrast between the intermediate and folded state due to either the used dye pairs (Förster distance of 70 Å in the case of 175C-298C Atto565-Alexa647), the close proximity of the attached dyes (in the case of N-C interface 52C-175C) or the broad FRET histogram distributions measured in three-color smFRET (in the case of DM-MBP 52PrK-175C-298C).

**Table 2**: Förster distance and correction factors used for various combinations of the dye-pairs in this study. *Indicates the corrected $\gamma$ used in the case of Alexa647 lifetime changes, see Supplementary Figure 3G.

| | Förster distance, $R_0$ (Å) | Detection correction factor, $\gamma$ | Spectral cross-talk, $\alpha$ | Direct acceptor excitation, $\delta$ |
|---|---|---|---|---|
| Atto532-Aexa647 (Two color) | 62 | 0.50/0.37* | 0.03 | 0.07 |
| Atto532-Atto647N (Two color) | 59 | 0.59 | 0.02 | 0.06 |
| Atto488-Atto565 (Three-color /Two-color) | 63 | 0.40/0.44 | 0.07 | 0.05 |
| Atto488-Alexa647 (Three-color /Two-color) | 53 | 0.20/0.25 | 0.01 | 0.01 |
| Atto565-Alexa647 (Three-color /Two-color) | 70 | 0.43/0.50 | 0.14 | 0.13 |

**Table 3:** Comparison of distances from accessible volume (AV) simulations and two-color FRET experiments performed with all the three double cysteine domain mutants of DM-MBP and labeled with the Atto532-Alexa647 dye-pair. Distances derived from AV simulations are the FRET averaged distances ($<R_{DA}>_E$) and extracted using the FPS software as explained in the materials and methods (Kalinin et al., 2012). Experimental distances were estimated using the photon distribution analysis (PDA) (Antonik et al., 2006). PDA fitting was applied on the *PR* histogram obtained from burst-wise binned raw photon counts. The Monto Carlo method was used to simulate the Gaussian distance distributions for the D-A separation with one or two populations in the *PR* histogram. d1 and d2 are the two derived inter-dye distances for the peaks of the two populations with their respective fractions $f_1$ and $f_2$ and widths (standard deviations) $\sigma_1$ and $\sigma_2$.

|  | A-V (Å) | Experimental | | | | | |
|---|---|---|---|---|---|---|---|
|  |  | d1(Å) | $\sigma_1$(Å) | $f_1$ | d2(Å) | $\sigma_2$(Å) | $f_2$ |
| Native-NTD (52C-298C) | 51.3 | 48.3 | 3.45 | 0.87 | 55.4 | 10 | 0.13 |
| Native-CTD (175C-298C) | 54.5 | 49.6 | 3.6 | 0.80 | 57.6 | 10 | 0.20 |
| Native-N-C interface (52C-175C) | 40.0 | 44.8 | 2.8 | 1.00 | - |  | - |
| Denatured-NTD (52C-298C) | - | - | - | - | 83.8 | 17 | 1.00 |
| Denatured-CTD (175C-298C) | - | - | - | - | 80.9 | 13 | 1.00 |
| Denatured-N-C interface (52C-175C) | - | 49.1 | 4.5 | 0.07 | 74.9 | 5.0 | 0.93 |

**Table 4:** Donor (Atto532) fluorescence lifetimes ($\tau_1$ and $\tau_2$), their fractions ($f_1$ and $f_2$), the corresponding calculated FRET efficiencies ($E_1$ and $E_2$) and the distance estimations (*d1* and *d2*) from a biexponential fit of the fluorescence decay for different double cysteine mutants of DM-MBP labeled with Atto532-Alexa647 (see Material and Methods). For comparison, the lifetime fractions from native state and under denaturing condition are given. Refolding was performed at the mentioned concentrations of GuHCl. All fits are shown as Supplementary figure 5.

| NTD (52C-298C) | | $\tau_1$ (ns) | $f_1$ | $E_1$ | d1 (Å) | $\tau_2$ (ns) | $f_2$ | $E_2$ | d2 (Å) | $\chi^2_{red}$ |
|---|---|---|---|---|---|---|---|---|---|---|
| Native | | 0.55 | 0.84 | 0.84 | 46.6 | 2.06 | 0.16 | 0.42 | 65.2 | 1.50 |
| Refolding in GuHCl (M) | 0.1 | 0.58 | 0.79 | 0.83 | 47.2 | 2.14 | 0.21 | 0.40 | 66.1 | 1.61 |
| | 0.2 | 0.81 | 0.51 | 0.77 | 50.5 | 2.79 | 0.49 | 0.22 | 76.2 | 1.58 |
| | 0.3 | 0.51 | 0.58 | 0.85 | 45.9 | 2.59 | 0.42 | 0.28 | 72.5 | 1.20 |
| | 0.5 | 0.68 | 0.30 | 0.80 | 48.6 | 2.97 | 0.70 | 0.17 | 80.3 | 1.03 |
| | 0.9 | 0.47 | 0.38 | 0.86 | 45.3 | 2.92 | 0.62 | 0.18 | 79.2 | 1.07 |
| | 2 | 0.85 | 0.22 | 0.76 | 51.0 | 3.24 | 0.78 | 0.09 | 89.5 | 1.60 |
| CTD (175C-298C) | | $\tau_1$ (ns) | $f_1$ | $E_1$ | d1 (Å) | $\tau_2$ (ns) | $f_2$ | $E_2$ | d2 (Å) | |
| Native | | 0.53 | 0.76 | 0.85 | 46.3 | 1.95 | 0.24 | 0.45 | 63.7 | 1.38 |
| Refolding in GuHCl (M) | 0.1 | 0.47 | 0.66 | 0.86 | 45.2 | 2.36 | 0.34 | 0.34 | 69.1 | 2.03 |
| | 0.2 | 0.53 | 0.64 | 0.85 | 46.3 | 2.90 | 0.36 | 0.19 | 78.7 | 1.75 |
| | 0.3 | 0.63 | 0.61 | 0.82 | 47.9 | 3.01 | 0.39 | 0.16 | 81.4 | 1.65 |
| | 0.5 | 0.58 | 0.51 | 0.83 | 47.1 | 3.18 | 0.49 | 0.11 | 86.9 | 1.94 |
| | 0.9 | 0.56 | 0.45 | 0.84 | 46.8 | 3.15 | 0.55 | 0.12 | 85.8 | 1.95 |
| | 2 | 0.91 | 0.24 | 0.74 | 51.8 | 3.14 | 0.76 | 0.12 | 85.7 | 1.68 |
| N-C (52C-175C) | | $\tau_1$ (ns) | $f_1$ | $E_1$ | d1 (Å) | $\tau_2$ (ns) | $f_2$ | $E_2$ | d2 (Å) | |
| Native | | 0.13 | 0.92 | 0.96 | 36.1 | 2.29 | 0.08 | 0.36 | 68.1 | 1.21 |
| Refolding in GuHCl (M) | 0.1 | 0.49 | 0.74 | 0.86 | 45.7 | 2.38 | 0.26 | 0.33 | 69.3 | 1.63 |
| | 0.2 | 0.62 | 0.69 | 0.82 | 47.7 | 2.62 | 0.31 | 0.27 | 73.0 | 1.85 |
| | 0.3 | 0.54 | 0.59 | 0.84 | 46.5 | 2.61 | 0.41 | 0.27 | 72.9 | 1.74 |
| | 0.5 | 0.80 | 0.56 | 0.77 | 50.3 | 2.58 | 0.44 | 0.28 | 72.4 | 1.62 |
| | 0.9 | 0.35 | 0.59 | 0.90 | 42.8 | 2.49 | 0.41 | 0.30 | 71.0 | 1.44 |
| | 2 | 0.65 | 0.23 | 0.81 | 48.3 | 2.95 | 0.77 | 0.17 | 79.9 | 1.71 |

**Table 5:** Results from the dynamic photon distribution analysis for all the three two-color DM-MBP constructs labeled with Atto532-Alexa647. Data for refolding measurements performed in 0.2, 0.3, 0.5, and 0.9 M GuHCl were analyzed. By fixing the two inter-dye distances for the folded and unfolded states ($R_F$ and $R_U$ respectively) estimated from the lifetime analysis (Table 4), the rates for interconversion ($k_{U \to F}$ and $k_{F \to U}$) between the states were extracted. Widths were kept at a constant fraction of the distance $R$ (0.07 $R$) (Kalinin, Sisamakis, et al., 2010). The relaxation time, $\tau_R$, was calculated as $1/(k_{U \to F} + k_{F \to U})$. See Material and Methods for a description of the analysis. * In the marked cases, distances used in the dynamic PDA analysis were extracted from the FRET efficiencies measured in the intermediate GuHCl titrations or from the native state because the fluorescence lifetime values at very high and very low FRET efficiencies can be difficult to determine correctly (e.g. due to donor and acceptor quenching and/or donor only/acceptor blinking events). Errors are the 95% confidence intervals obtained from the Jacobian fit. See Supplementary Figure 7.

| NTD (52C-298C) | | $R_F$ (Å) | $k_{F \to U}$ (ms$^{-1}$) | $R_U$ (Å) | $k_{U \to F}$ (ms$^{-1}$) | $\tau_R$ (µs) |
|---|---|---|---|---|---|---|
| | 0.1 | 47.2 | 0.12±0.03 | 66.1 | 0.72±0.19 | 1190±287 |
| Refolding | 0.2 | 45.9* | 1.03±0.10 | 76.2 | 3.23±0.27 | 234±15 |
| in GuHCl | 0.3 | 45.9 | 1.38±0.12 | 72.5 | 2.13±0.22 | 284±19 |
| (M) | 0.5 | 48.6 | 2.08±0.21 | 80.3 | 0.89±0.10 | 336±26 |
| | 0.9 | 45.3 | 1.91±0.17 | 79.2 | 0.63±0.07 | 393±28 |
| CTD (175C-298C) | | $R_F$ (Å) | $k_{F \to U}$ (ms$^{-1}$) | $R_U$ (Å) | $k_{U \to F}$ (ms$^{-1}$) | $\tau_R$ (µs) |
| | 0.1 | 45.2 | 1.19±0.09 | 69.1 | 2.19±0.17 | 295±16 |
| Refolding | 0.2 | 46.3 | 1.02±0.11 | 78.7 | 1.98±0.37 | 333±43 |
| in GuHCl | 0.3 | 47.9 | 1.17±0.11 | 81.4 | 2.76±0.22 | 254±15 |
| (M) | 0.5 | 47.1 | 2.31±0.22 | 81.4* | 3.00±0.28 | 188±12 |
| | 0.9 | 46.8 | 3.95±0.29 | 81.4* | 1.60±0.12 | 180±10 |
| N-C (52C-175C) | | $R_F$ (Å) | $k_{F \to U}$ (ms$^{-1}$) | $R_U$ (Å) | $k_{U \to F}$ (ms$^{-1}$) | $\tau_R$ (µs) |
| | 0.1 | 44.8* | 0.13±0.04 | 69.3 | 0.62±0.16 | 1333±307 |
| Refolding | 0.2 | 44.8* | 0.39±0.05 | 73.0 | 3.34±0.59 | 268±43 |
| in GuHCl | 0.3 | 46.5 | 0.77±0.07 | 72.9 | 3.85±0.33 | 216±16 |
| (M) | 0.5 | 44.8* | 0.55±0.08 | 72.4 | 3.75±0.49 | 232±27 |
| | 0.9 | 46.5* | 1.70±0.23 | 71.0 | 3.51±0.55 | 191±22 |

# Main Figures
## Figure 1

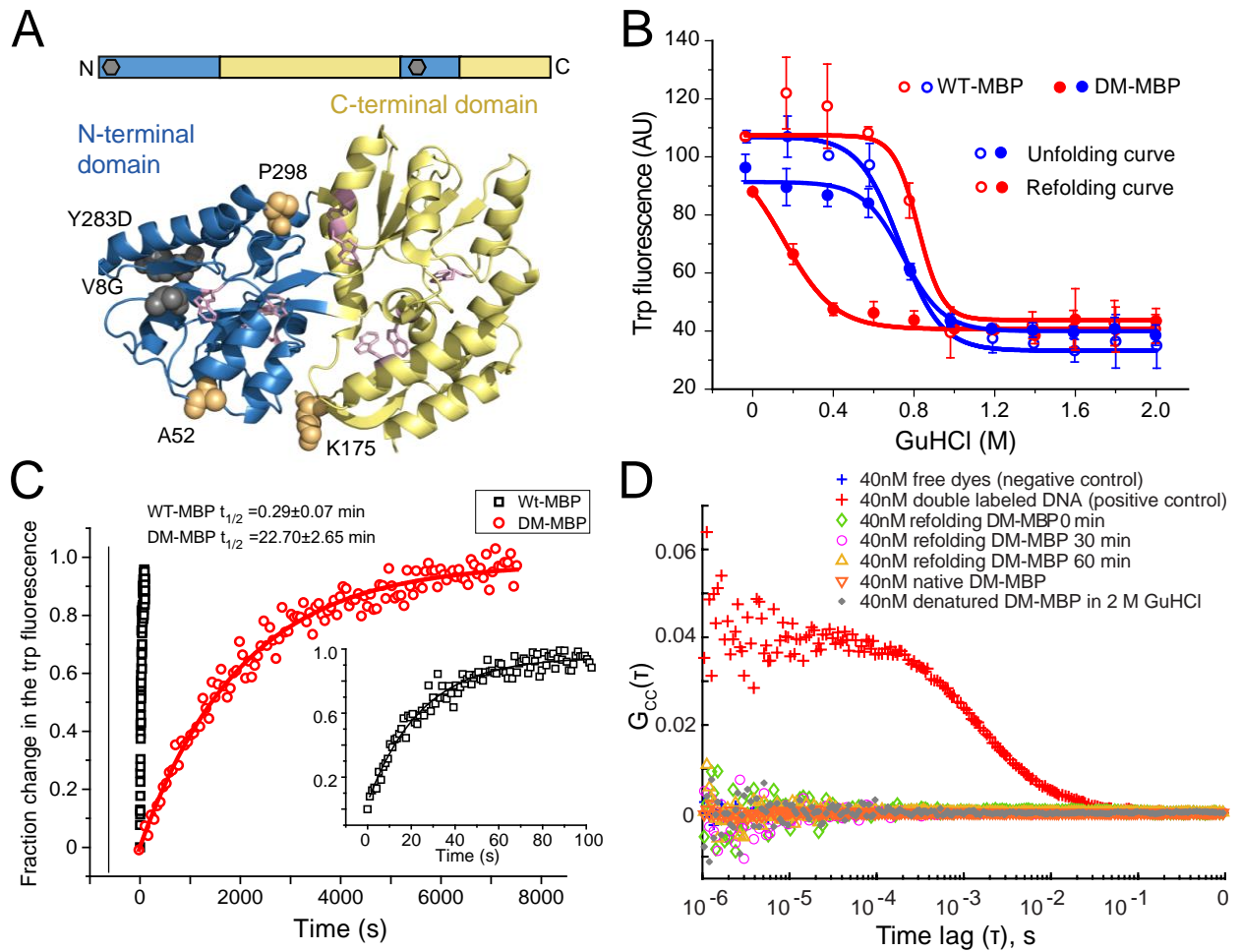

**Figure 1. DM-MBP unfolding / refolding**

**(A)** A ribbon structure of the maltose binding protein (MBP) (PDB ID:1OMP) showing the N- (NTD) and C-terminal domain (CTD) in blue and yellow respectively. The upper schematic represents the discontinuity in MBP sequence for the NTD and CTD where the positions of the mutations V8G and Y283D are depicted as grey hexagons. The residues involved in the two folding mutations and the three labeling positions A52, K175, P298 for coupling the fluorescent dyes are indicated via a space filling model in dark grey and orange respectively. Eight tryptophan side-chains are highlighted as a stick model in pink.

**(B)** Equilibrium unfolding and refolding experiments in different GuHCl concentrations of WT (open circles) and DM-MBP (closed circles) measured with tryptophan fluorescence. To record the unfolding curve (blue), steady state tryptophan fluorescence of ~40 nM native MBP was measured after 20 hr in the respective GuHCl concentration at 22°C. For the refolding curve (red), 2 µM of MBP was first denatured in 3 M GuHCl for 1 hr at 50°C, diluted 50-fold and incubated for 3-4 hrs at the indicated final GuHCl concentrations for before measuring the steady state tryptophan fluorescence. The error was estimated from at least from three independent titrations. Each curve was fitted with a Boltzmann function (see Materials and Methods). Estimated error is a standard deviation.

(**C**) The kinetics of WT (black squares) and DM-MBP (red open circles) refolding monitored by the increase in tryptophan fluorescence. The initial fluorescence at time t=0 was subtracted from the subsequent data points. 3 µM MBP was denatured in 3 M GuHCl for 1 hr at 50°C before being diluted 75-fold in Buffer A to start the refolding reaction (at t=0, the final concentrations were ~40 nM of MBP and 40 mM of GuHCl). Data were fitted with a single exponential function. The fit to the WT-MBP refolding is shown for initial 100 s in the inset for the clarity. The presented curve is a single measurement representative of three independent measurements. The half-life ($t_{1/2}$) for the refolding process are reported in the figure.

(**D**) Fluorescence cross-correlation spectroscopy (FCCS) measurements of a mixture of 20 nM Atto532 labeled DM-MBP (A52C) and 20 nM Alexa647 labeled DM-MBP undergoing refolding during 0 min, 30 min and after 60 min (green, magenta and yellow curves, respectively). 500-1000 nM DM-MBP (A52C) was first denatured in 3 M GuHCl. The sample was then diluted first serially in 3 M GuHCl before the final dilution to achieve the nanomolar concentrations of the labeled protein. The FCCS amplitude was used to analyze the presence of small oligomers in the sample. Atto532 and Atto647N labeled DNA served as a positive control for a cross-correlation signal (red curve). Freely diffusing Atto532 and Atto655 dyes were measured as a negative control with no cross-correlation amplitude visible (blue curve). For completeness, we also performed the FCCS analysis under native (orange curve) and denaturing (grey curve) conditions.

# Figure 2

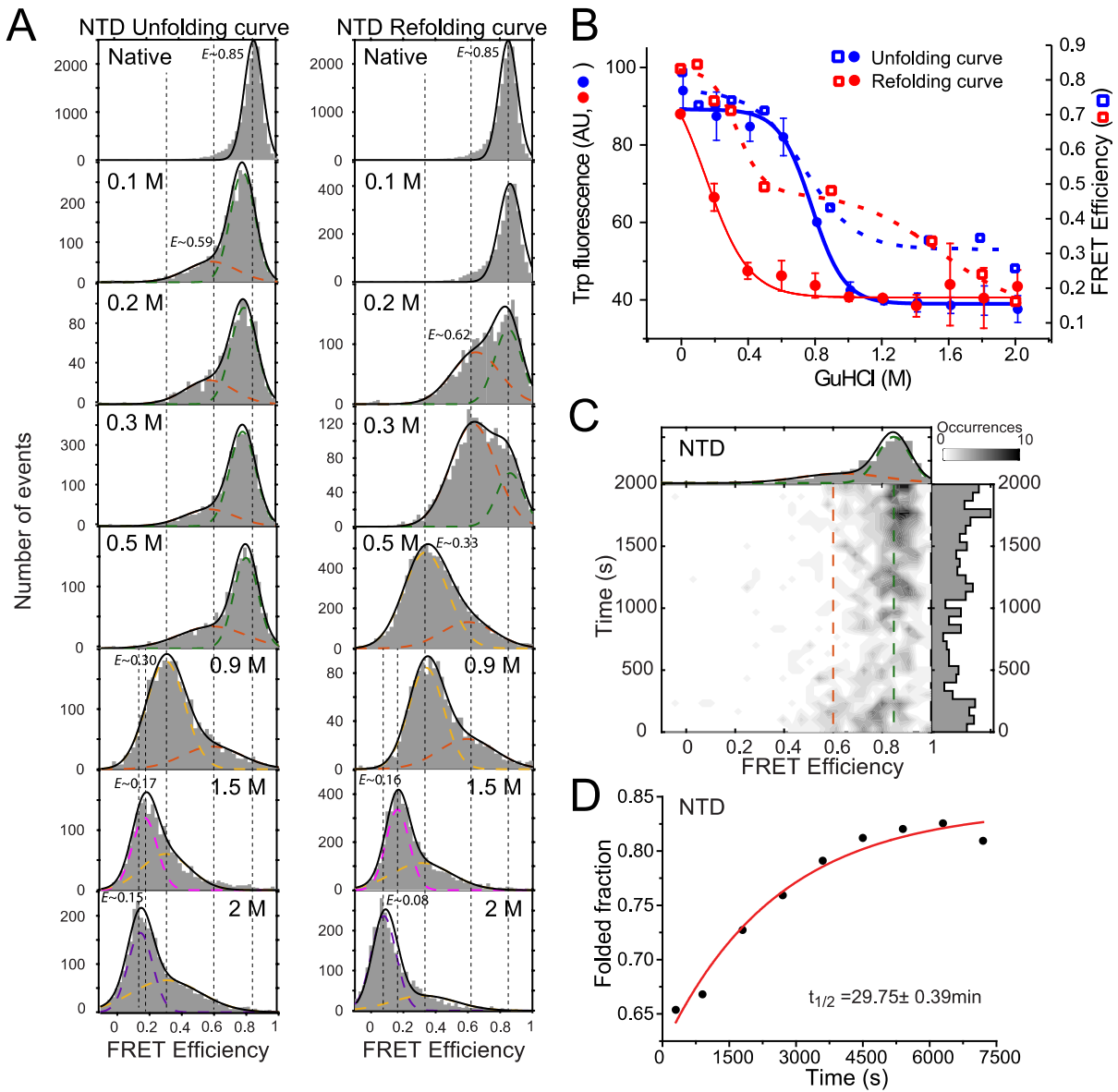

**Figure 2. Equilibrium unfolding-refolding two-color smFRET measurements on the NTD reveals a refolding intermediate population**

**(A)** SmFRET histograms and corresponding Gaussian fits for DM-MBP unfolding and refolding titrations at the indicated denaturant concentrations. Each underlying population is highlighted with a dotted line. The native or folded state is shown in green, intermediate states are shown in orange and in yellow, and the completely unfolded state is shown in violet (2 M GuHCl). The equilibrium unfolding curve is obtained by performing two-color smFRET measurements on native DM-MBP (52-298) in buffer containing different GuHCl concentrations. To measure the refolding curve, 500-1000 nM DM-MBP (52-298) was first denatured in 3 M GuHCl for 1 hr at 50°C. The sample was then diluted first serially in 3 M GuHCl at room temperature before the final 50- fold dilution to achieve the picomolar concentrations of labeled protein in buffer or buffer/GuHCl mixture at the indicated final GuHCl concentrations and refolding commences immediately. Protein refolds completely within an hour when refolding was performed in 0.1 M GuHCl. The histogram does not change in the other measurements throughout the course of the measurement times of 3-4 hr.

**(B)** To dissect out the heterogeneity in the ensemble curves measured in Figure 1B, average FRET efficiencies obtained from (A) were compared to the equilibrium unfolding and refolding curve measured with tryptophan fluorescence at the same denaturant titrations. The tryptophan curves (circles) and average FRET efficiencies (squares) for unfolding and refolding titrations are depicted in blue and red respectively. Ensemble curves are taken from Figure (1B). Average FRET efficiency refolding and unfolding curves were fitted with a single and double Boltzmann function respectively.

**(C)** Kinetics of NTD refolding was analyzed for the initial 2000 s. Refolding was induced after 30-fold dilution of 3 M GuHCl denatured NTD. A Gaussian fit to the FRET histogram shows the presence of an intermediate population with a FRET efficiency of 0.6 in addition to the folded fraction at 0.85 FRET efficiency.

**(D)** The refolding rate [half-life ($t_{1/2}$)] for the reaction in **(C)** was calculated by assaying the increase in the folded fraction (0.85 FRET efficiency) as a function of time. The red curve is a mono-exponential fit to the data (shown in black). Also see Table 1. The presented curve is a single measurement representative of three independent measurements.

Figure 3

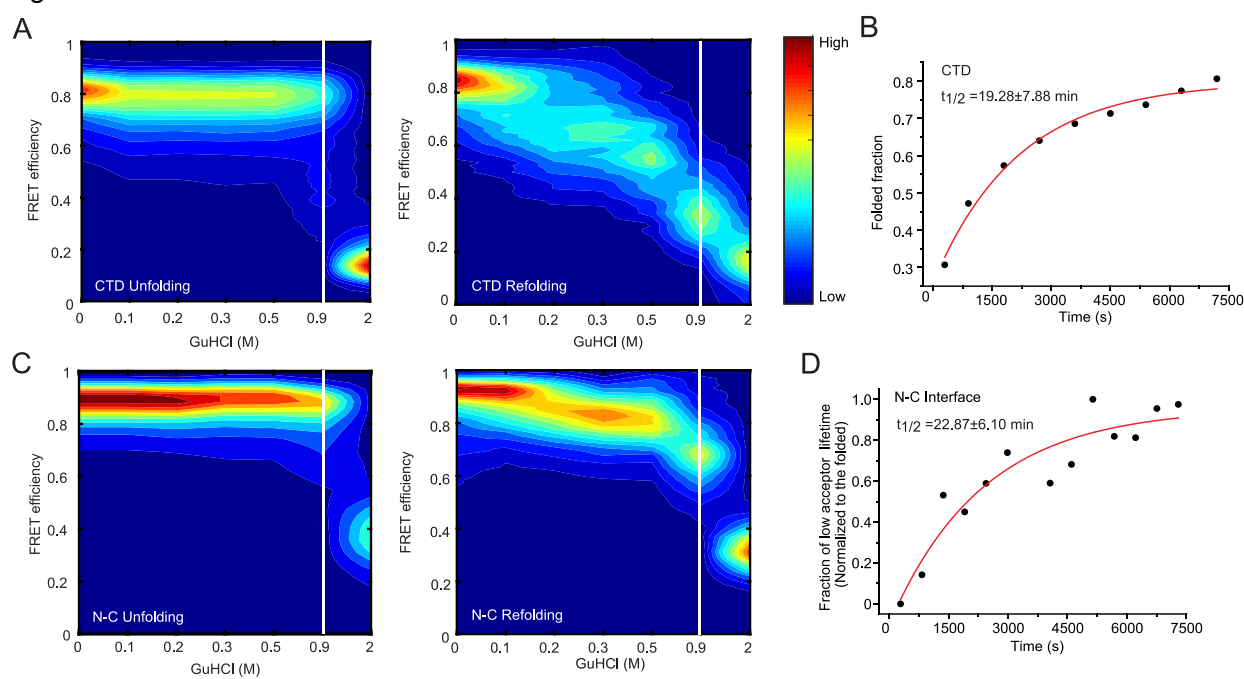

**Figure 3. Evidence for an intermediate state under kinetic and equilibrium refolding conditions**

**(A-B)** Equilibrium and kinetic analyses of CTD unfolding and refolding. **(A)** 2D-plots of FRET efficiency versus denaturant concentration are presented in a waterfall scheme for equilibrium unfolding (left) and refolding (right) measurements. For denaturant concentrations between ~ 0.2 and 0.5 GuHCl, an intermediate state with a FRET efficiency of 0.6 is visible. **(B)** A kinetic analysis of FRET histograms for CTD refolding. Refolding was induced after 30-fold dilution of 3 M GuHCl denatured protein. The red curve is a mono-exponential fit to the data (black). The half-life ($t_{1/2}$) for the refolding process is 19.28±7.88 min (see Table 1).

**(C-D)** Equilibrium and kinetic analyses of N-C interface. **(C)** 2D-plots of FRET efficiency versus denaturant concentration are presented in a waterfall scheme for equilibrium unfolding (left) and refolding (right) measurements. Similar to the NTD and CTD, the N-C interface also shows an intermediate state that is unique to the refolding process. **(D)** A kinetic analysis for N-C interface refolding was assayed from smFRET experiments. Refolding was induced after 30-fold dilution of 3 M GuHCl denatured protein. Refolding was monitored via the increase in the low acceptor lifetime (see Supplementary Figure 3). The red curve is a mono-exponential fit to the data (black). The half-life ($t_{1/2}$) for the refolding process is 22.87±6.10 min (see Table 1 and Table S2). The presented curve is a single measurement representative of three independent measurements.

Figure 4

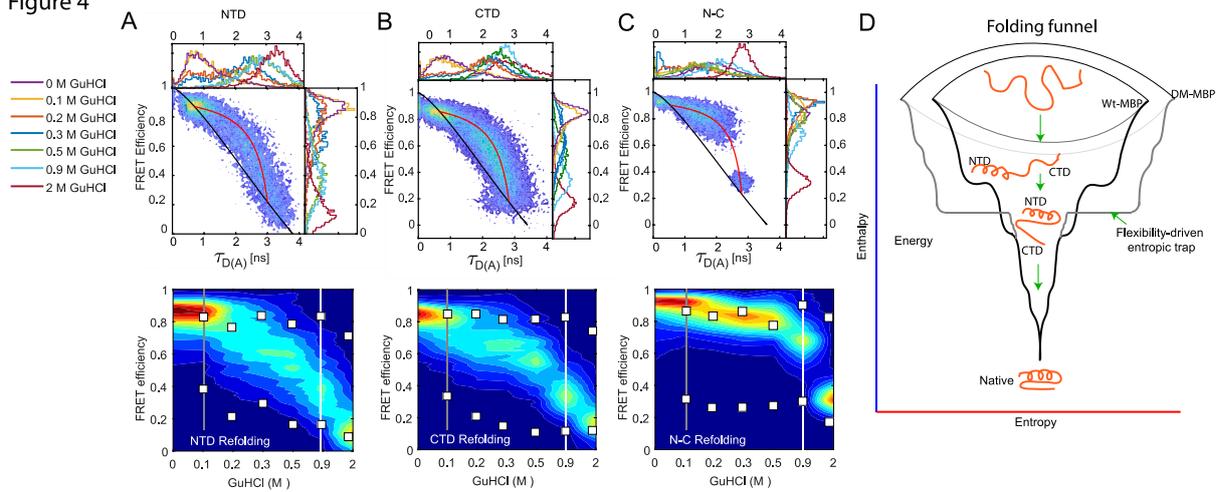

**Figure 4. FRET efficiency versus donor lifetime analysis shows that the conformational search is the cause for then entropic trap**

**(A-C)** Upper panels: 2D-plots of FRET efficiency vs donor lifetime in presence of an acceptor ($\tau_{D(A)}$) ($E$-$\tau$ plot), for NTD equilibrium refolding measurements shown in the right panel of Figure 2A (A), for CTD equilibrium refolding measurements shown in the right panel Figure 3A (B) and for N-C interface experiments shown in the right panel of Figure 3C (C). For comparison, all measurements are superimposed. The 1-D projections of the different GuHCl measurements, color coded according to the legend, are shown above and to the right of the 2D plots. The ideal relationship between FRET efficiency and donor lifetime is shown in black (static-FRET line) for a molecule that remains in a single FRET state during the observation time. Dynamic interconversion on micro- to millisecond time-scales between the folded and unfolded FRET states leads to a deviation in the $E$-$\tau$ plot, as shown in red (dynamic-FRET line). See Materials and Methods. For simplicity, the dynamic FRET red line is only shown for the 0.9 M GuHCl refolding experiment.

Lower panels: A waterfall scheme comparing the burst averaged FRET efficiency versus denaturant concentration for NTD refolding (A), CTD refolding (B) and formation of the N-C interface (C). The burst-averaged FRET efficiencies were compared with the FRET efficiencies obtained from the two donor lifetime components determined from a fit to the photons detected from all bursts. For each denaturant concentration, the donor lifetime derived FRET efficiencies represent two conformational states which are interconverting during refolding, shown as filled squares in white. A white line separates the 0.9 M GuHCl measurement from the higher denaturant concentrations, as there is a continuous distance increase in the unfolded state between 1 to 2 M GuHCl (see Table 4). Note that the donor lifetime derived FRET efficiency values for the unfolded state continuously decreases with increasing denaturant concentration. Grey line indicates the folded conformation of NTD and N-C interface, whereas CTD is still yet to be fold when refolding was performed in 0.1 M GuHCl.

**(D)** A schematic of the folding funnel describing the origin of entropic barrier resulting in intermediate state during the refolding process. Vertical axis represents the enthalpic energy and horizontal axis represents the configurational entropy. Dark funnel is for Wt-MBP and grey funnel for DM-MBP. In case of Wt-MBP, NTD folds faster with a guided by the hydrophobic interaction, which then is followed by CTD folding. In case of DM-MBP, due to the disrupted hydrophobic nucleus, NTD folding is limited by the high configurational entropy leading to slow folding. This step represents the higher degree of native contacts needed to stabilize the NTD folding than typically required by Wt-MBP to compensate for the loss of binding energy due to double mutations. As soon as the NTD finds its folding competent conformation in DM-MBP, CTD folds faster and follows the rest of the folding funnel as for Wt-MBP. Presence of a flexibility in the DM-MBP refolding experimentally demonstrated with FRET efficiency versus donor lifetime analysis highlights the extended region in the horizontal plane creating an entropic barrier towards the folded native state.

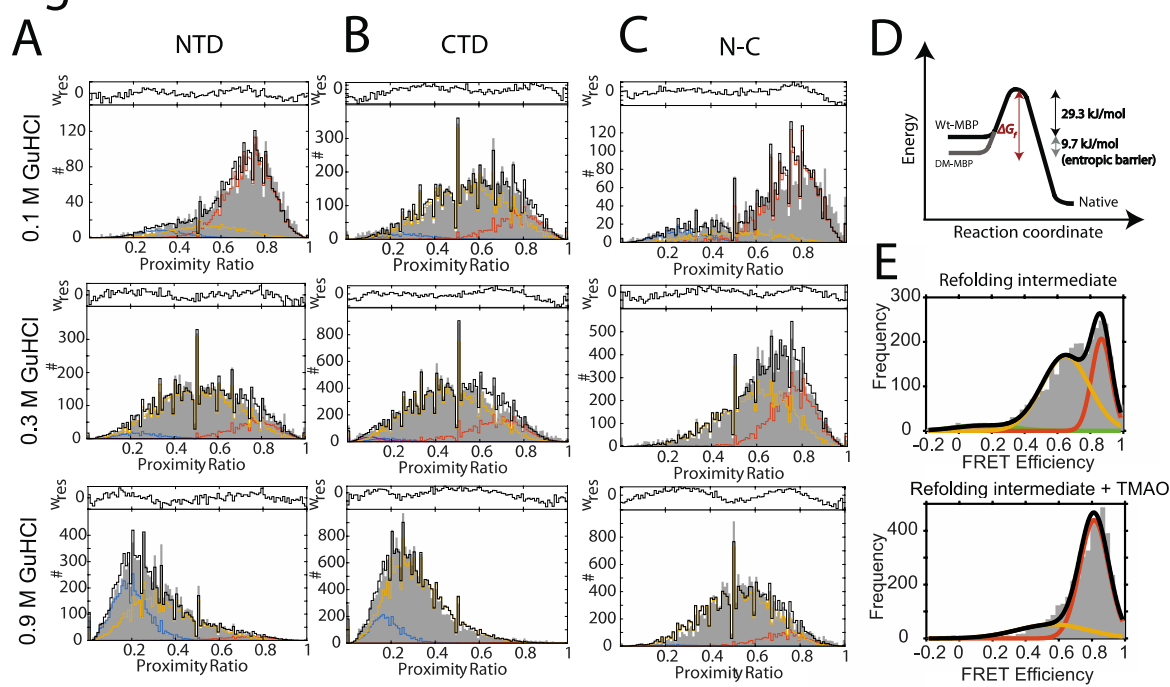

**Figure 5. Quantification of the energy barrier in an entropic trap**

**(A-C)** A dynamic PDA analysis for NTD refolding (**A**), CTD refolding (**B**) and for the N-C interface (**C**) in 0.1 M GuHCl (upper panels), 0.3 M GuHCl (middle panels), and 0.9 M GuHCl (lower panels). The initial 2000 s of the refolding measurements were analyzed in the case of refolding in 0.1 M GuHCl for all the three constructs. In grey are the proximity ratio (*PR*) histograms with 1 ms binning. The dynamic PDA fitting is highlighted as a black outline in the histogram. In the dynamic PDA fit, the folded population is highlighted in red, the unfolded population in blue, and yellow represents the contribution of the interconverting species to the *PR*. Photons were binned into histograms with 0.5, 1 and 1.5 ms binning and were fitted globally to quantify the transition rates. See Supplementary Figure 6B-D and Table 5 for all the titrations.

**(D)** A schematic showing the Gibbs free energy versus reaction coordinate, where an additional energy barrier of 10.55 kJ/mol is imposed by the flexible-entropically trapped intermediate on DM-MBP folding.

**(E)** Upper panel: A smFRET histogram for NTD refolding in 0.2 M GuHCl, where a significant fraction of the intermediate population is observed.

Lower panel: A smFRET histogram for NTD refolding in 0.2 M GuHCl where 500 mM trimethylamine-N-oxide has been added. The trimethylamine-N-oxide acts as a molecular chaperone by confining the conformational space available to the protein through stabilization of the solvent shell around the protein. In the case of DM-MBP, this allows the protein to overcome the entropic barrier, leading to faster folding and an increase in the native conformation in 0.2 M GuHCl.

# Figure 6

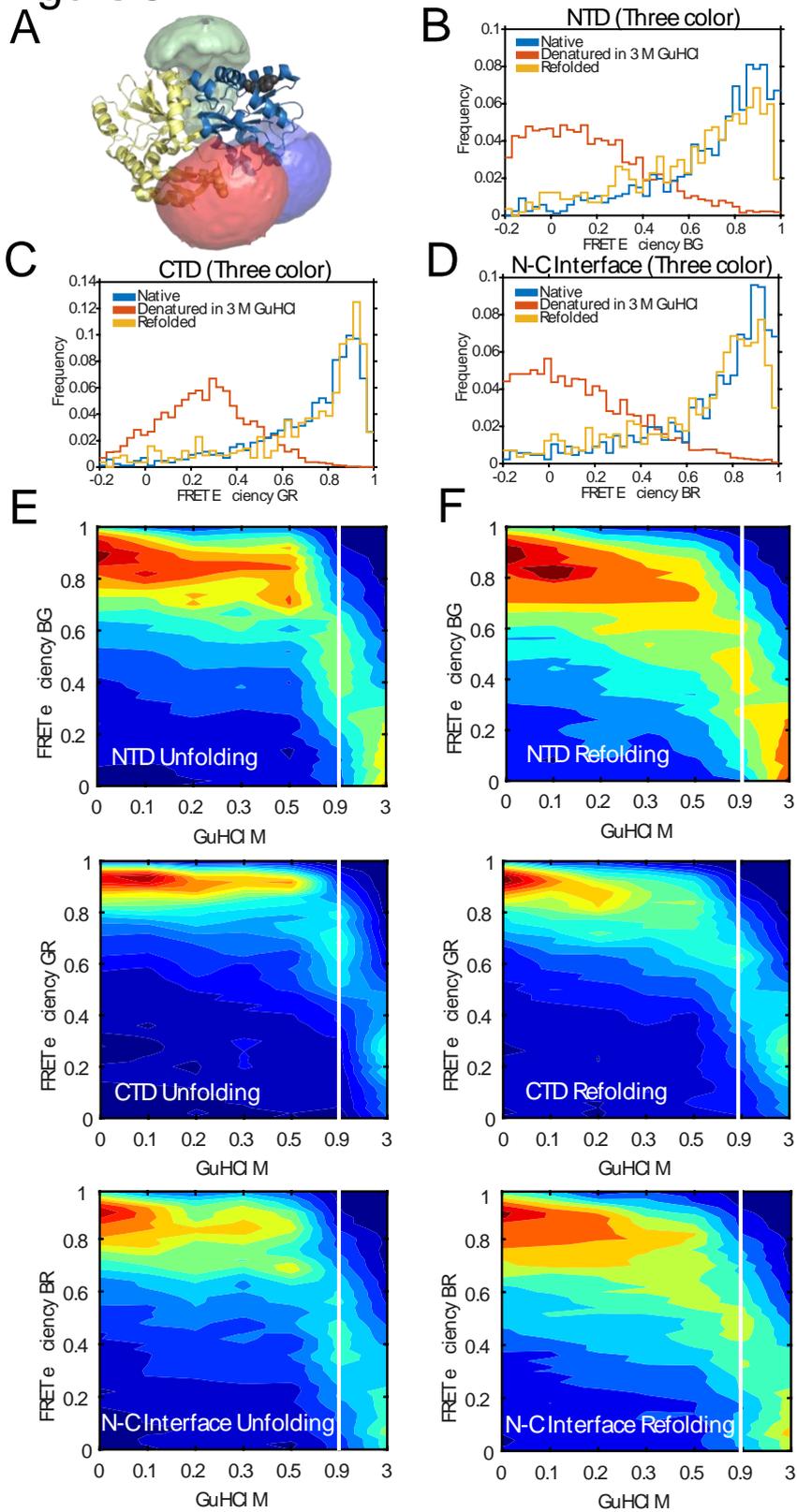

**Figure 6. Three-color smFRET demonstrates the co-existence of an intermediate population**

**(A)** Accessible volume calculations for triply-labeled DM-MBP on the MBP structure (PDB ID:1OMP) at the labeling positions A52, K175 and P298 for Atto488, Atto565 and Alexa647 dyes respectively.

**(B)-(D)** SmFRET histograms from the three-color FRET experiments for the three FRET pairs GR, monitoring the NTD (**B**), BG monitoring the CTD (**C**) and BR for the N-C interface (**D**). SmFRET histograms are compared for the native conformation (blue), denatured state (in 3 M GuHCl in orange) and refolded conformation (yellow).

**(E)-(F)** 2D-plot of FRET efficiency versus GuHCl concentration, presented as waterfall plots to visualize the conformational changes during equilibrium unfolding (**E**) and refolding **(F)** of triple-labeled DM-MBP. The upper panels show three color-BG FRET histograms indicative of the NTD, middle panels show three color-GR FRET histograms indicative of CTD and lower panel show three color-BR FRET histograms for N-C interface. For the equilibrium unfolding experiments, three-color FRET measurements were performed on native triple-labeled DM-MBP diluted into different GuHCl concentrations. For the refolding measurements, first, DM-MBP was denatured in 3 M GuHCl for 1 hr at 50°C. The sample was allowed to refold by diluting the sample to indicated final GuHCl concentrations. The white line separates the 0.9 M GuHCl measurement from the higher denaturant concentrations, where the average separation in the unfolded state continues increase from 1 to 2 M GuHCl.

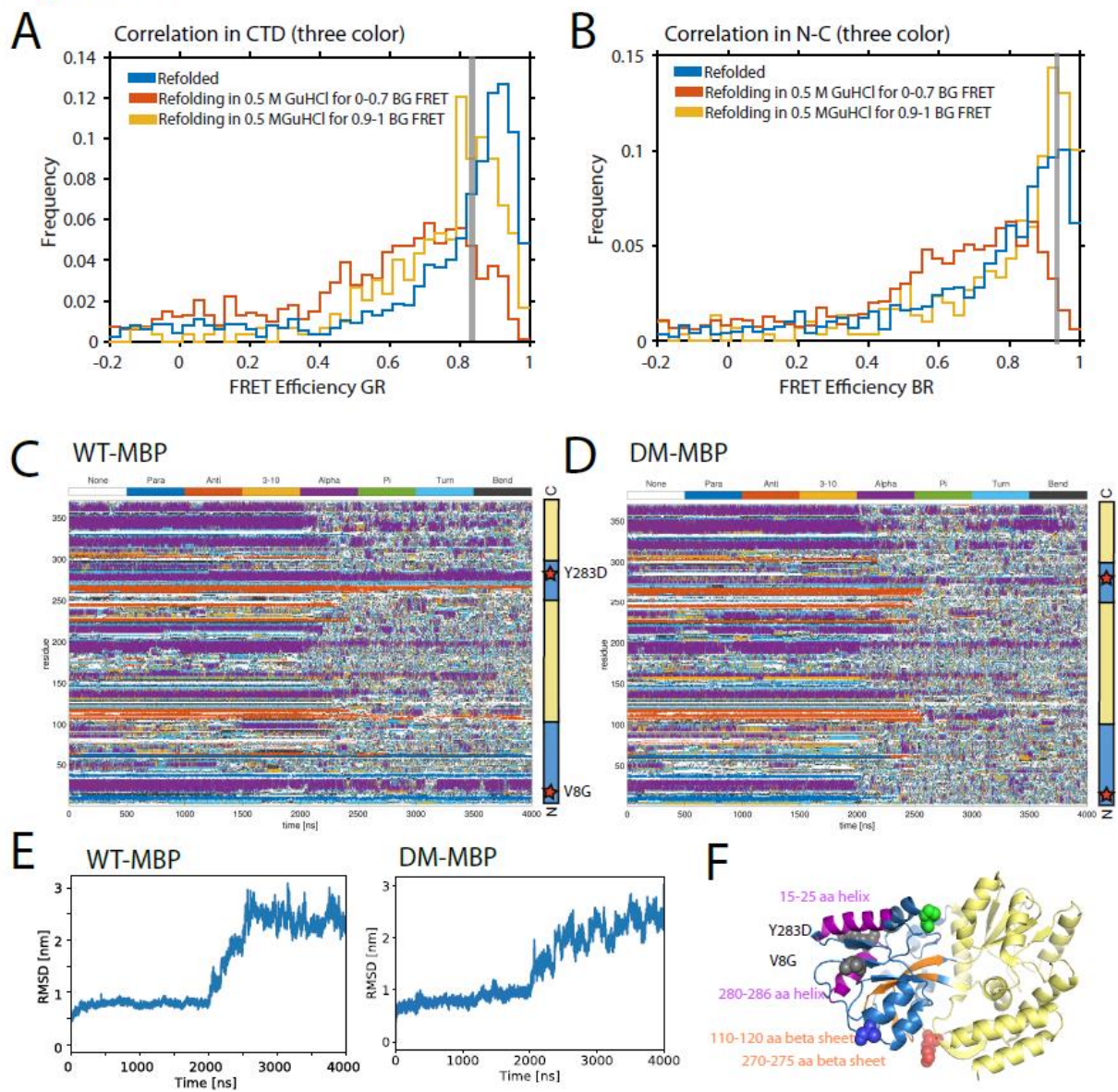

**Figure 7. Correlative refolding analysis from three-color smFRET and unfolding MD simulations**

(**A-B**) A comparison of the three color smFRET histograms for molecules in the intermediate state (0< $E_{BG}$ < 0.7, orange) and in the native-like conformation ($E_{BG}$ > 0.9, yellow), as determined from the FRET efficiency in the NTD. The three color-BR smFRET histograms for the CTD (**A**) and the three-color-BR for the N-C interface (**B**) are shown. Measurements of triple-labelled DM-MPB refolding in 0.5 M GuHCl were used for the analysis. All the three FRET efficiencies are available simultaneously from the same molecule. For comparison, the respective smFRET histograms for the refolded state are shown in blue. Grey line shows that the CTD is not folded yet while N-C interface has folded when molecules were selected for folded state for NTD.

**(C-D)** DSSP (Definition of Secondary Structure of Proteins) based secondary structure annotation plot for MD simulations based on PDB ID:1OMP were performed on WT-MBP (**C**) and DM-MBP (**D**) during temperature-induced unfolding. The plot highlights the various secondary structures elements present in amino acid sequence from N to C terminus as a function of simulation time. The simulation was performed at 400 K for the initial 2 µs to reach equilibrium and then temperature was than increased to 450 K for another 2 µs to induce unfolding. Color scheme: random coil is shown in white, parallel beta sheets in blue, anti-parallel beta sheets in orange, 3-10 helices in yellow, alpha helices in purple, Pi helices in green, beta turns in sky blue and bends in black. Predominantly found alpha helices and anti-parallel beta sheets secondary structures can be seen throughout the MBP structure. The locations of the double mutations present in DM-MBP, V8G and Y283D, are depicted in right panel in the MBP sequence.

**(E)** RMSD (Root mean square deviation) plot for MD simulations performed during temperature induced unfolding from panel C on WT-MBP (left panel) and from panel D on DM-MBP for (right panel)**.** RMSD is calculated for backbone atoms over a period of MD simulation with respect to the initial structure.

**(F)** Secondary structures are highlighted on the MBP structure that are preserved in the MD simulations performed on WT-MBP after heating at 450 K as shown in panel C. Alpha helices are shown in purple, anti-parallel beta sheets are shown in orange. Double mutations in dark grey and three dye labels are shown in their respective color codes used for three-color smFRET.

Figure 8

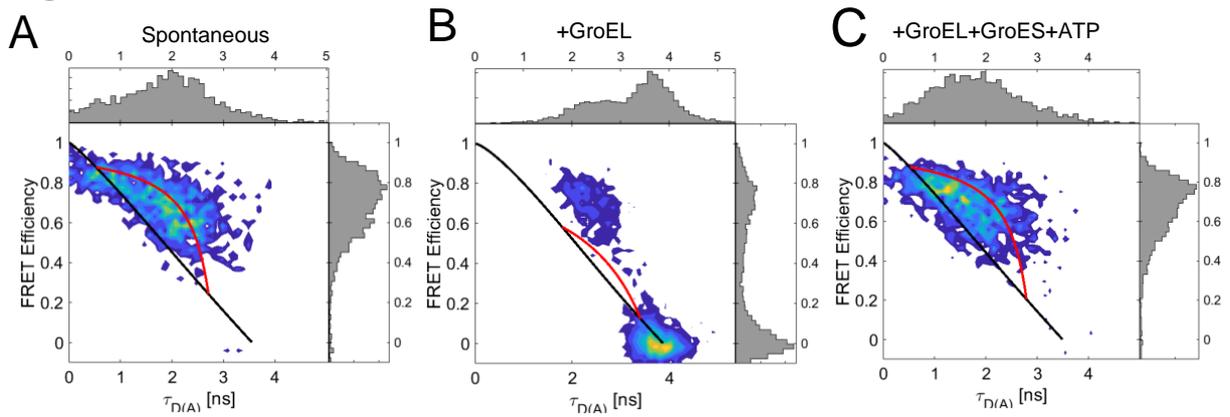

**Figure 8. The influence of GroEL and GroEL\ES\ATP mediated confinement on the folding landscape of DM-MBP**

**(A-C)** 2D FRET efficiency vs donor lifetime ($\tau_{D(A)}$) histograms ($E$-$\tau$ plot) of the NTD during refolding in 0.1 M GuHCl. The refolding of the NTD is shown for (A) spontaneous folding, (B) in presence of 3 µM GroEL, and (C) upon the addition of 3 µM GroEL, 6 µM GroES and 2 mM ATP. Only the initial 10 minutes of all the three measurements are shown. For GroEL and GroEL\ES\ATP measurements, the added components were incorporated in the dilution buffer. The 1-D projections are shown above and to the right of the 2D plots. The ideal relationship between FRET efficiency and donor lifetime is shown in black (static-FRET line) for a molecule that remains in a single FRET state during the observation time. Dynamic interconversion on micro- to millisecond time-scales between the folded and unfolded FRET states leads to a deviation in the $E$-$\tau$ plot, as shown in red (dynamic-FRET line). The two donor fluorescence lifetimes were extracted from the measurements to construct the dynamic-FRET line. See Materials and Methods for details.

# Supplementary document

## A dynamic intermediate state limits the folding rate of a discontinuous two-domain protein

Ganesh Agam, Anders Barth, and Don C. Lamb

# Supplementary Figures

## Supplementary Figure 1

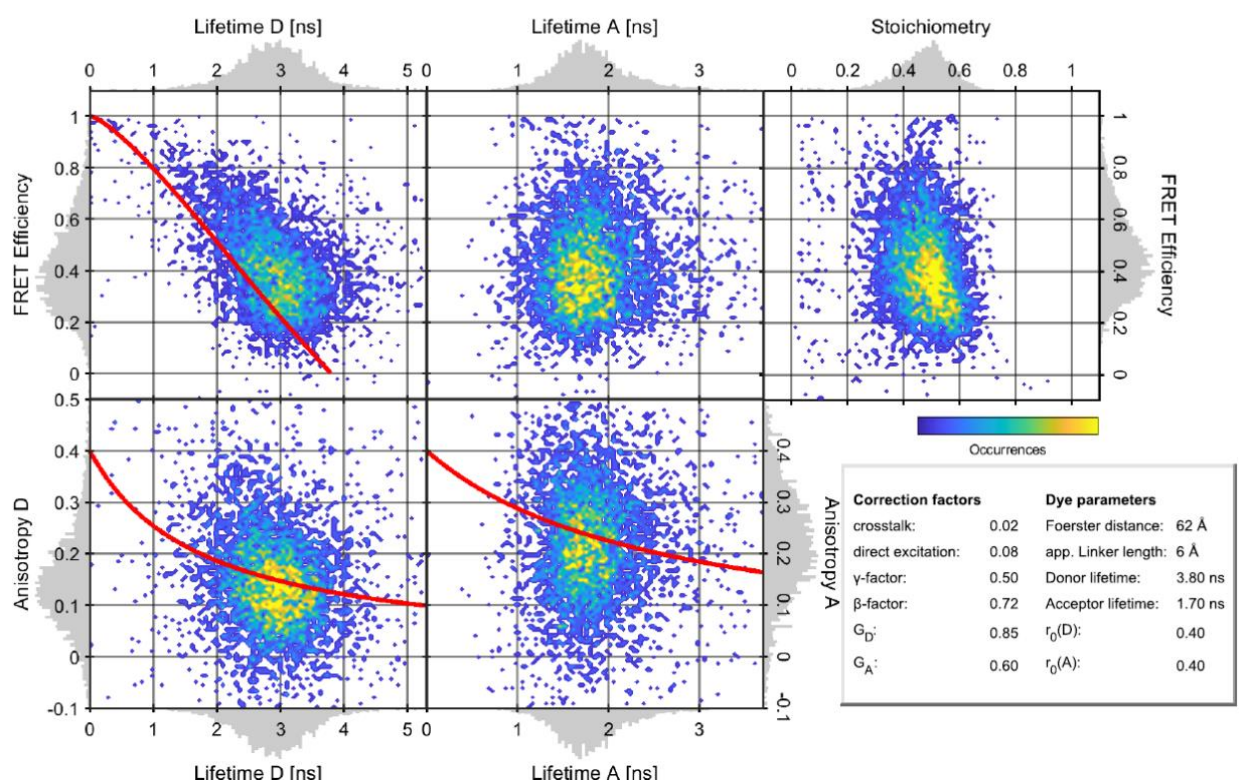

**Supplementary Figure 1. Data analysis employed with multi-parameter fluorescence detection and pulsed interleaved excitation (MFD-PIE) scheme used for all two-color smFRET measurements.**

A representative all-in-one plot for two-color smFRET measurements for the NTD refolding performed in 0.9 M GuHCl. Each burst represents a single molecule Upper right panel shows, stoichiometry (S) of ~0.5 (typical for double labeled molecules) and FRET efficiency for the analyzed molecules. Upper left panels show the relationship between FRET efficiency and burst-wise donor lifetime (Lifetime D in ns) and acceptor lifetime (Lifetime A in ns). The ideal relationship between the FRET efficiency and donor lifetime is highlighted by the red line.
Lower panels depict the burst-wise anisotropy values for donor (D) and acceptor (A) fluorophores with their respective lifetimes. Red lines are fits to the Perrin equation. The applied correction factors are shows on lower right panel (see Table 2).

# Supplementary Figure 2

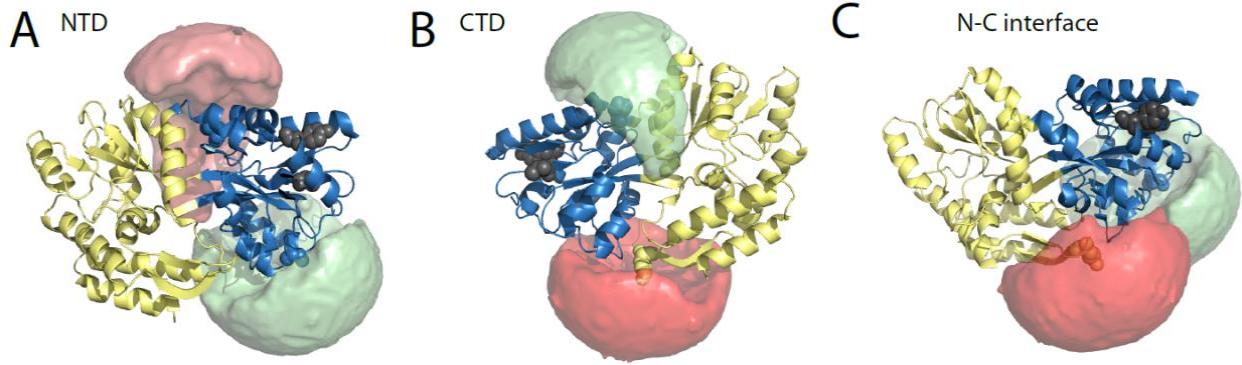

**Supplementary Figure 2. Accessible volume simulations of NTD, CTD and N-C interface (A-C)** Accessible volume calculations on the MBP structure (PDB ID:1OMP) using Atto532 and Alexa647 dyes. Calculations were performed for only one dye combination of double labels for used constructs of NTD, CTD and N-C interface. For NTD, positions A52 (Atto532) and P298 (Alexa647) **(A)**, for CTD K175 (Alexa647) and P298 (Atto532) **(B)** and for N-C interface A52 (Atto532) and K175 (Alexa647) **(C)** were used for calculations with particular dye mentioned in the parentheses. See the Materials and Methods, and Table 3 for the details.

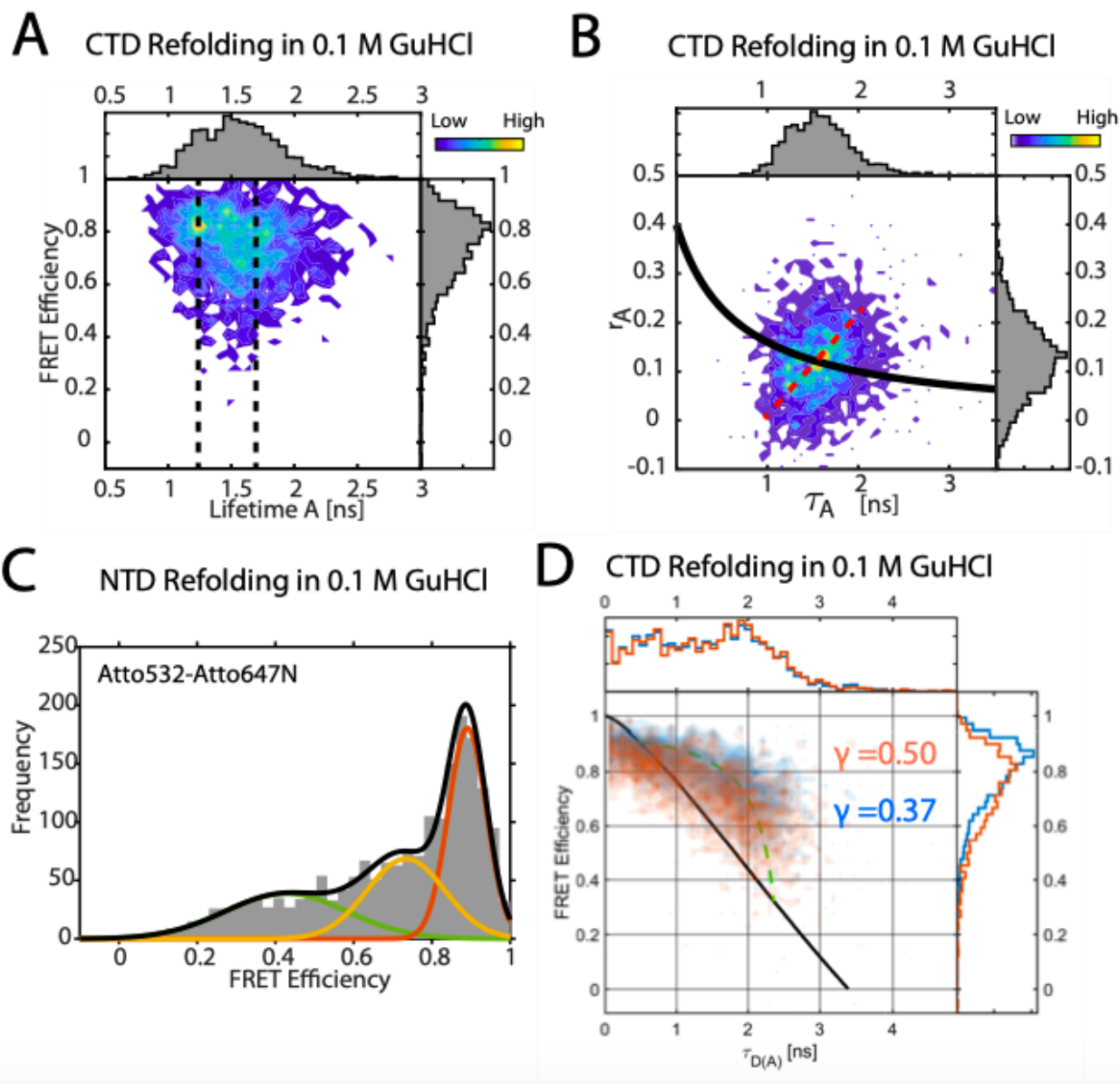

**Supplementary Figure 3. Photophysical property of Alexa647 observed in MBP refolding studies**

**(A)** A 2D-plot of FRET efficiency versus acceptor lifetime ($\tau_A$) during refolding of the CTD. A decrease in the fluorescence lifetime of Alexa647 from 1.7 ns (for intermediate state) to 1.2 ns is visible for the folded protein.

**(B)** A 2D-plot of $r$, the steady state anisotropy versus acceptor lifetime ($\tau_A$) for the same refolding measurement as in panel A . The Perrin equation was fit to the steady state anisotropy and acceptor lifetime (black line) to find the rotational correlation time, $\rho$, and thereby analyze the rotation of a dye. Overall, a $\rho$ of 0.67 was found for the acceptor but a clear trend an increase in the steady state anisotropy with increasing lifetime is observable (dashed red line). Note that the steady state anisotropies are below 0.2.

**(C)** A smFRET histogram of Atto532-Atto647N labeled NTD from the first 3000 s of refolding. The intermediate state found when using Alexa647 was still preserved with Atto647N as an alternate acceptor.

**(D)** A 2D-plot of FRET efficiency versus $\tau_{D(A)}$ for CTD refolding. The detection correction factor, $\gamma$ was calculated for all the bursts as 0.5 for unquenched acceptor lifetime of 1.7 ns (shown in orange). This leads the acceptor quenched refolded population of ~0.8 FRET efficiency falling below the static-FRET line (black). When corrected to account for the quenched acceptor lifetime, with the corrected $\gamma$ of 0.37 (shown in blue), the refolded population (now at 0.85 FRET efficiency) correctly falls on the static-FRET line.

# Supplementary Figure 4

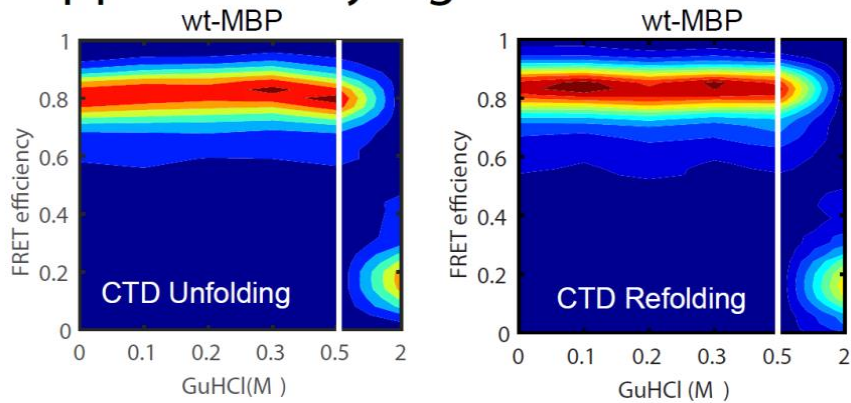

**Supplementary Figure 4. Equilibrium unfolding-refolding curves of wt-MBP shows no refolding intermediate**

2D-plots of FRET efficiency versus denaturant concentration are presented in a waterfall scheme for wt-MBP during unfolding (left) and refolding (right). Contrary to the refolding traces of the NTD, CTD, and the N-C interface for DM-MBP (Figures 2 and 3), no intermediate state is visible for wt-MBP refolding. A white line separates the measurements below and above 0.5 M GuHCl concentration, a concentration below which the intermediate state is noticeable during DM-MBP refolding.

# Supplementay Figure 5

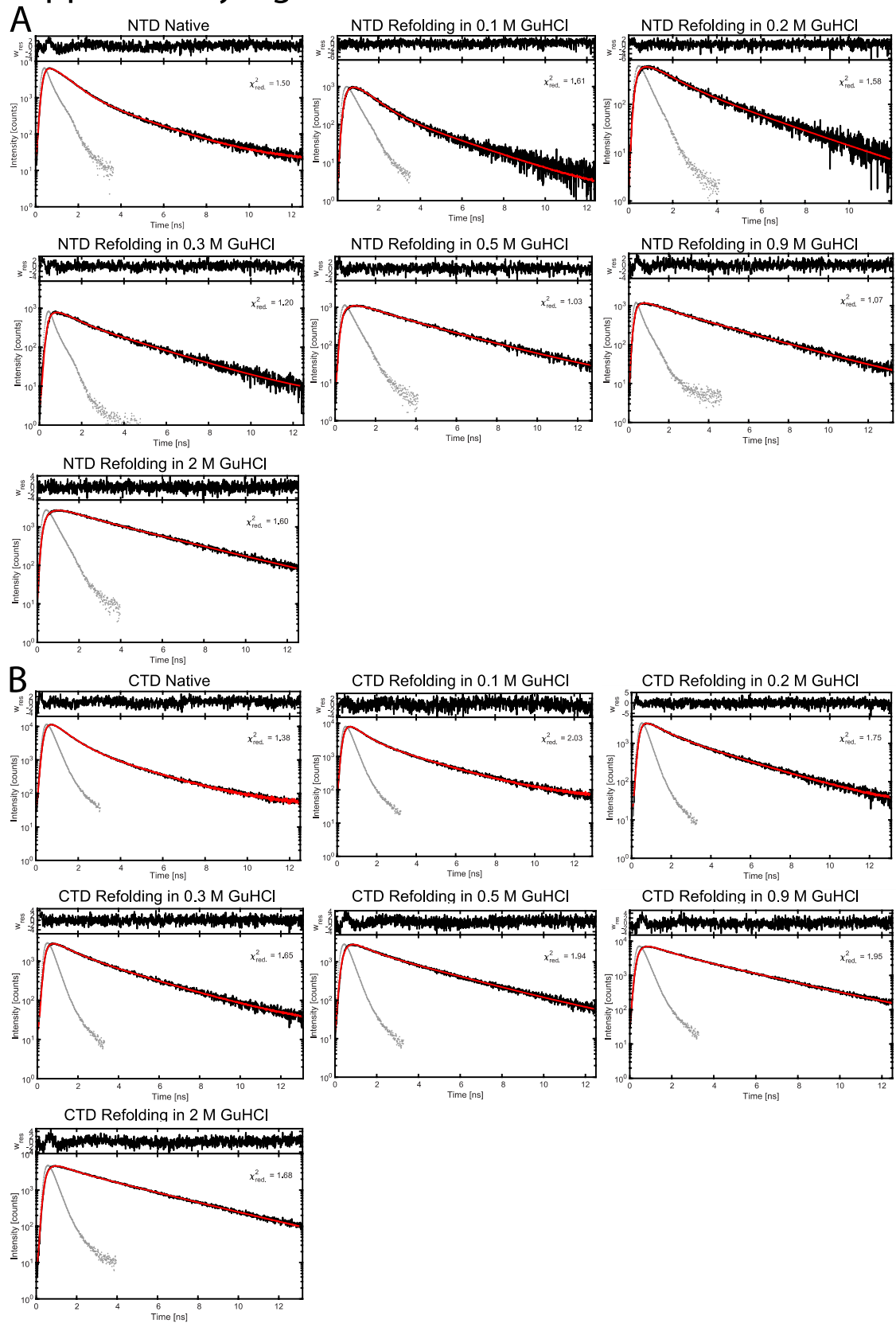

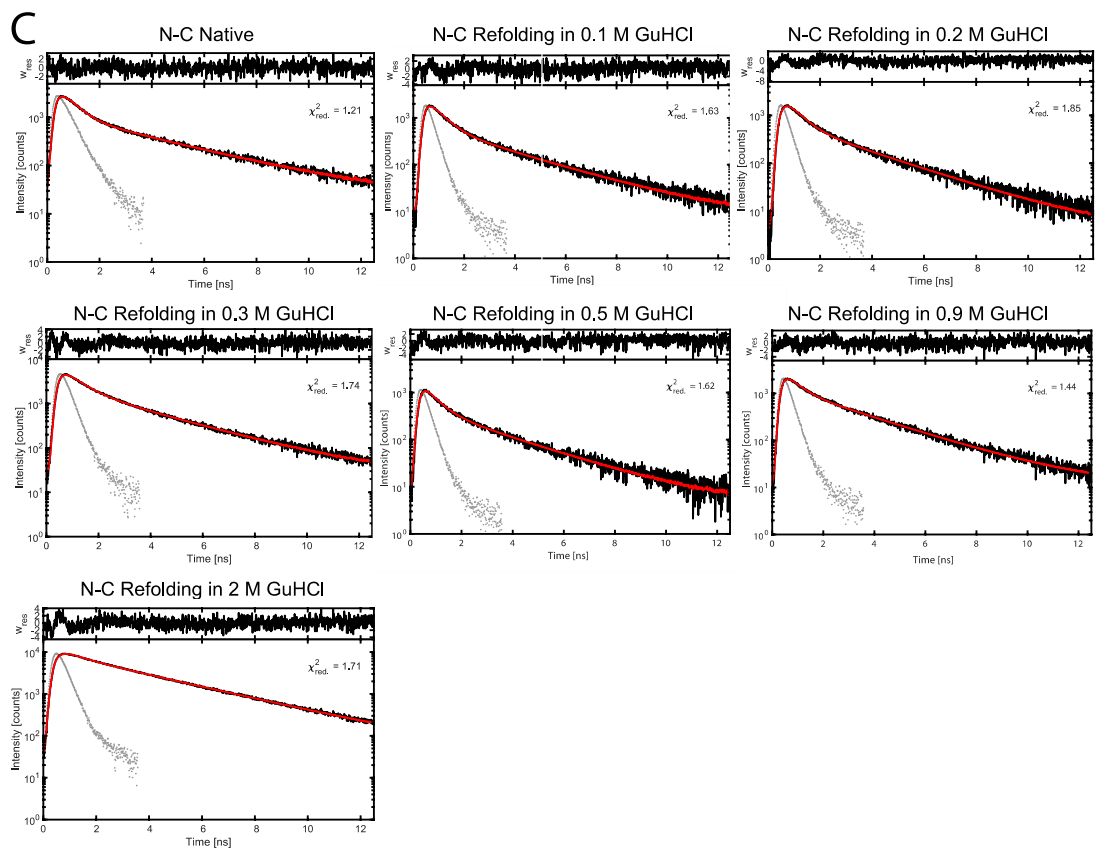

**Supplementary Figure 5. Analysis of the donor fluorescence lifetime for double-labeled molecules of DM-MBP from two-color FRET measurements on the NTD, CTD and N-C interface.**

**(A-C)** To extract the donor lifetimes defining the conformational states present in DM-MBP refolding, a donor fluorescence decay (black) was fitted with a bi-exponential function (red line) for NTD **(A)**, CTD **(B)** and N-C interface **(C)**. All the three constructs were labeled with Atto532 and Alexa647. The lifetime fit was performed on the intensity decay by a convolution with the instrument response function (grey). The quality of the fit model was judged by the $\chi^2_{red}$ value. The upper panel shows the weighted residuals from the fit. The donor lifetime values obtained from the fits are summarized in Table 4. These values were employed to carry out a FRET efficiency versus donor lifetime analysis as well as to analyze the conformational dynamics using the dynamic photon distribution analysis.

# Supplementay Figure 6

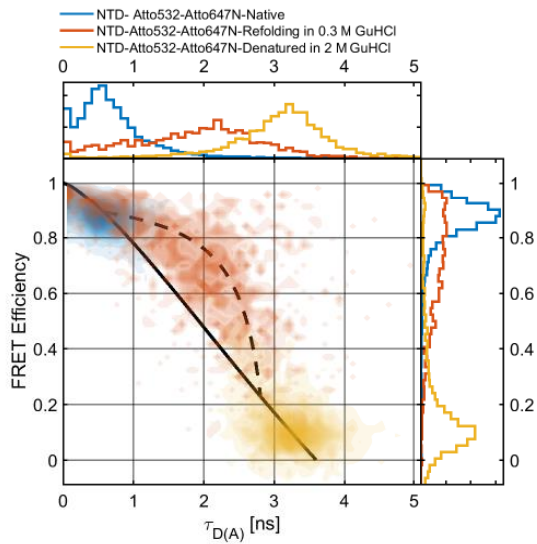

**Supplementary Figure 6. Conservation of conformational dynamics with different acceptor in DM-MBP folding.**

A 2D-plot of FRET efficiency versus donor lifetime in presence of an acceptor ($\tau_{D(A)}$) ($E$-$\tau$ plot) for the NTD construct labeled with Atto532 and Atto647N. FRET histograms for the native configuration (blue), refolding in 0.3 M GuHCl (orange) and in the denatured state (i.e. in 2.0 M GuHCl) (yellow) are shown. Note that the refolding NTD has a similar intermediate state possessing sub-millisecond dynamics as observed with Alexa647 as an acceptor.

# Supplementay Figure 7

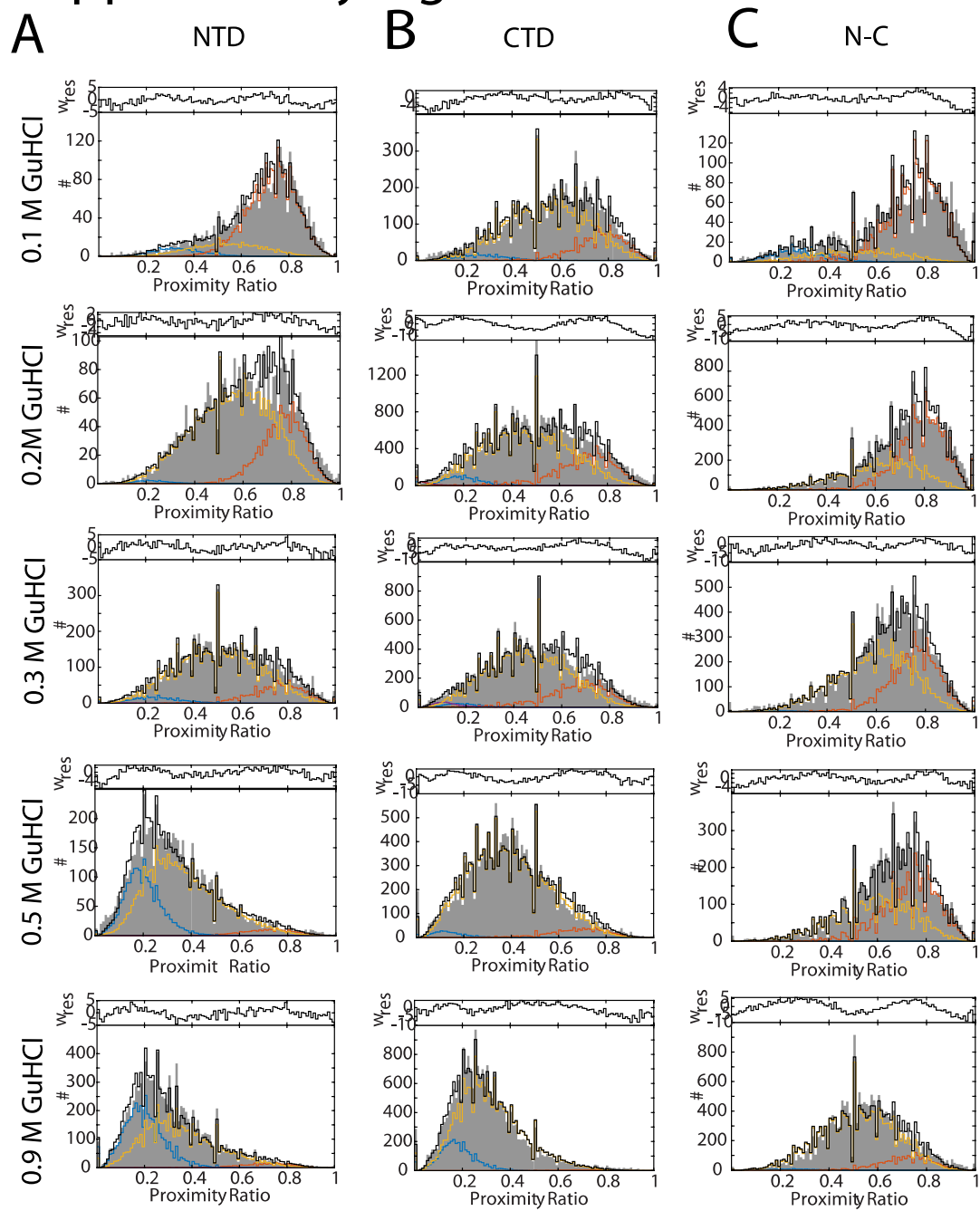

**Supplementary Figure 7. Conformational dynamics quantified using the dynamic photon distribution analysis**

**(A-C)** A dynamic PDA analysis for the NTD (**A**), CTD (**B**), and for the N-C (**C**) interface, labeled with Atto532 and Alexa647 are shown. Analyzed refolding measurements were performed in 0.1, 0.2, 0.3, 0.5, and 0.9 M GuHCl concentrations. The initial 2000 s of the refolding measurements were analyzed in the case of 0.1 M GuHCl refolding for all the three constructs. In grey are the proximity ratio histograms. The dynamic PDA fitting is highlighted with a black outline in the histogram. In dynamic PDA fitting, the folded population is highlighted in red, the unfolded state in blue, and yellow represents the contribution of the interconverting species between the folded and unfolded states to the histogram. A global fit was performed using 0.5, 1 and 1.5 ms binned proximity ratio histograms to quantify the transition rates (see Table 5). The 1 ms binned data are shown.

# Supplementary Figure 8

A

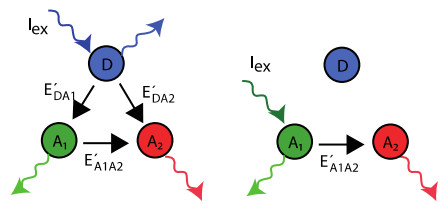

B

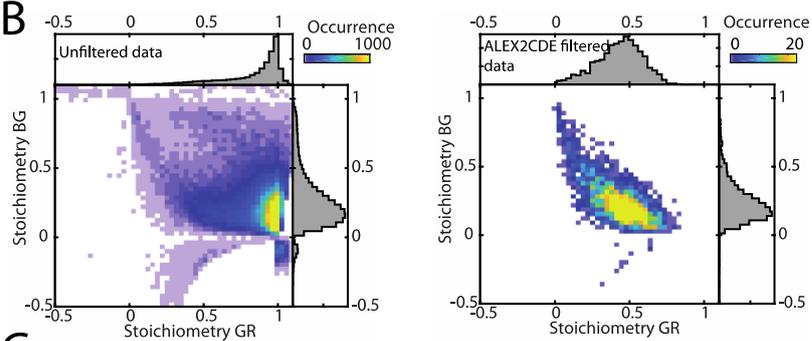

C

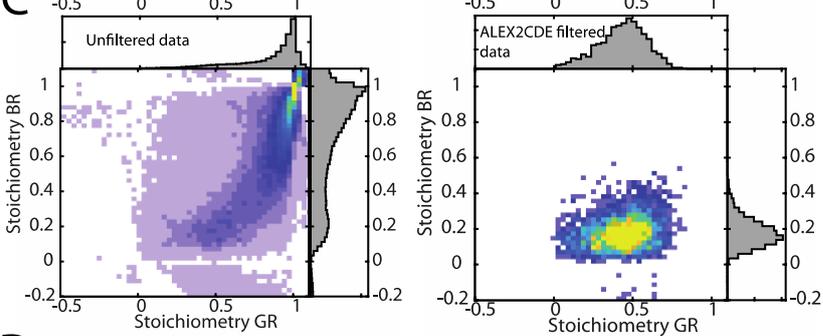

D

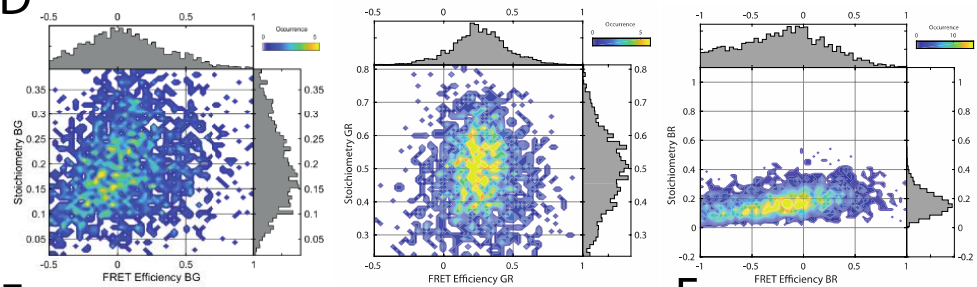

E

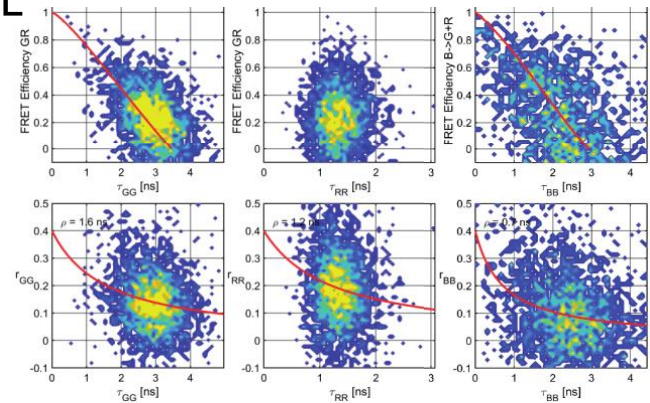

F

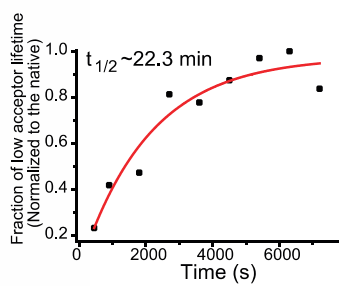

**Supplementary Figure 8. Three-color smFRET analysis of DM-MBP with MFD-PIE**

**(A)** Three-color smFRET scheme used in the study. In left panel, after exciting the blue dye (donor, D), energy can be transferred to the green dye (first acceptor, $A_1$) or the red dye (second acceptor, $A_2$). Furthermore, the excited green dye can transfer the energy to the red dye. The rate of energy transfer between the green and red dyes can be determined by directly exciting the green fluorophore with a green laser pulse, as shown in the right panel.

**(B)** Selection criterion used for analyzing three-dye labeled molecules. Left panel: A 2D plot of the BG stoichiometry versus GR stoichiometry for all the bursts. Right panel: The same 2D stoichiometry plot after applying the ALEX two-color 2CDE filter removing the potential photophysical artifacts of photobleaching and blinking. The remaining molecules with a stoichiometry of ~0.5 for GR and ~0.2 for BG were selected for further quantitative three-color smFRET analysis. In the case of three-color smFRET, the data were measured using MFD-PIE by alternating three laser pulses. See Material and Methods section for the details.

**(C)** Selection criterion used for analyzing three-dye labeled molecules. Left panel: A 2D plot of BR stoichiometry versus GR stoichiometry for all the bursts. Right panel: The same 2D stoichiometry plot after applying the ALEX two-color 2CDE filter removing the potential photophysical artifacts of photobleaching and blinking. The remaining molecules with a stoichiometry of ~0.5 for GR and ~0.2 for BR were selected for further quantitative three-color FRET analysis. In the case of three-color FRET, data were measured using MFD-PIE by alternating three laser pulses. See Material and Methods section for the details. Note that the selected molecules for further three-color FRET analysis must pass both the BG-GR and BR-GR selection criterion.

**(D)** Representative three-color FRET data of 3 M GuHCl denatured triple-labeled DM-MBP. Triple-labeled molecules were selected applying the criteria discussed in panels B and C. Left panel: The BG stoichiometry versus FRET efficiency BG is plotted. Middle panel: The GR stoichiometry versus FRET efficiency GR is plotted. Right panel: The BR stoichiometry versus FRET efficiency BR is plotted.

**(E)** Burst-wise lifetime and anisotropy plots of three-color FRET data for 3 M GuHCl denatured triple-labeled DM-MBP available with MFD-PIE. Upper panels, from left to right: The GR FRET efficiency versus green dye lifetime ($\tau_{GG}$), GR FRET efficiency versus red dye lifetime ($\tau_{RR}$) and B→G+R FRET efficiency versus blue dye lifetime ($\tau_{BB}$). Red lines represent the ideal relationship between FRET efficiency and donor (green/blue) lifetime for static FRET states. Lower panel, from left to right: The burst-wise anisotropy values for donor (blue and green) and acceptor (red) fluorophores versus their respective lifetimes are plotted. The red lines are the fits to the Perrin equation as explained in the Material and Methods section.

**(F)** A kinetic analysis of DM-MBP refolding of the three-color smFRET data assayed using the increase in the fraction of molecules with a low acceptor lifetime as explained in Supplementary Figure 3**D-E**. See Table 1 and Table S2.

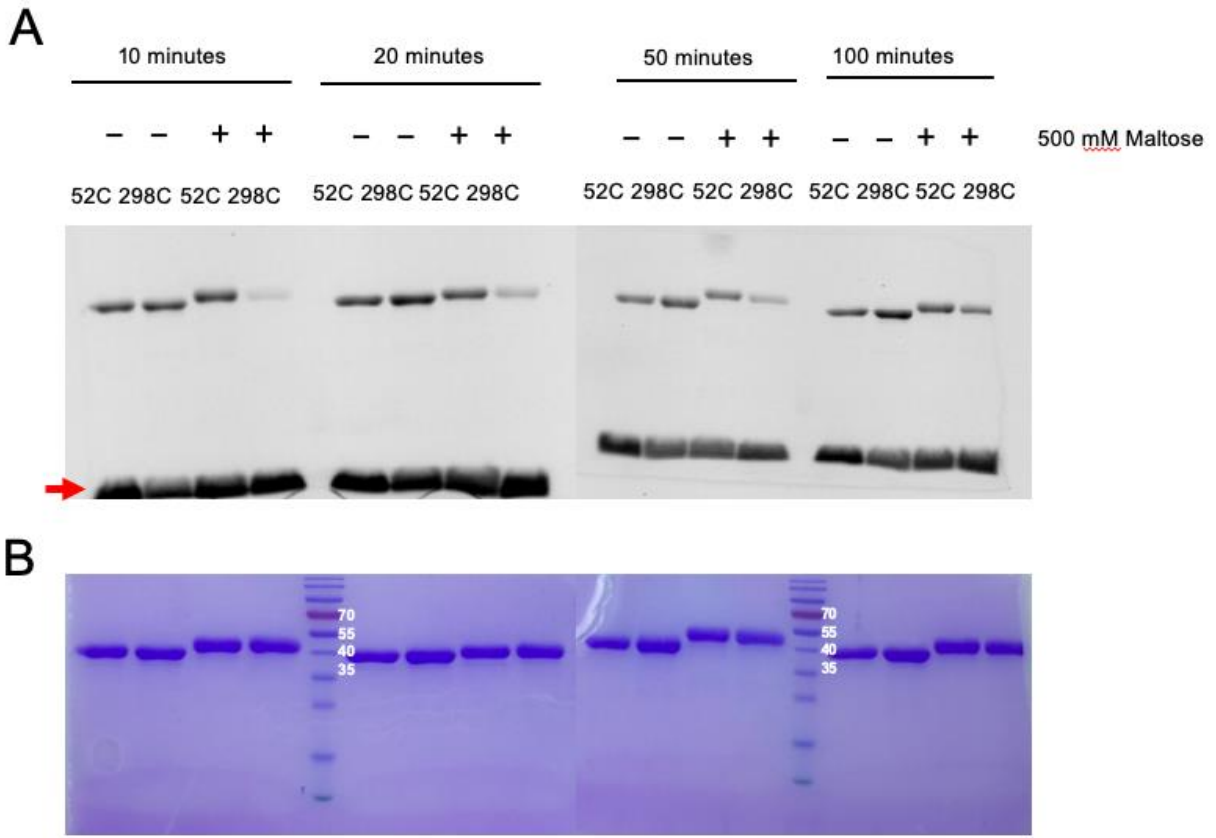

**Supplementary Figure 9: P298C position is inaccessible for cysteine-maleimide labelling when maltose is bound to DM-MBP.**

**(A)** Labeling of Atto565-maleimide conjugate to a single cysteine mutants A52C and P298C was monitored after 10 min, 20 min, 50 min and 100 minutes from the start of the reaction. Labeling was performed in the absence and the presence of 500 mM Maltose for the two mutants. The sample was loaded and run on a SDS-PAGE gel. The signal of the Atto565 fluorescence was collected by excitation with UV-light. Maltose binding to DM-MBP makes the P298C position less accessible to the labeling reaction, even after 50 minutes, whereas labeling to position A52C was evident within 10 minutes. A52C serves as a positive control for the labeling reaction at a solvent accessible position in presence of maltose. It also helps to rule out any viscosity effects arising from the dissolve 500 mM maltose on the labeling kinetics. Free unlabeled dye is indicated with a red arrow on the scanned gel. The presence of 500 mM maltose in the reaction slows down the migration of the SDS-denatured protein on the gel.

**(B)** A coomassie staining of the same SDS-PAGE gels after scanning for Atto565 emission. The molecular weights of the marker proteins are given in kDa. The amount of the protein loaded for all reactions were equal.

# Supplementary Figure 10

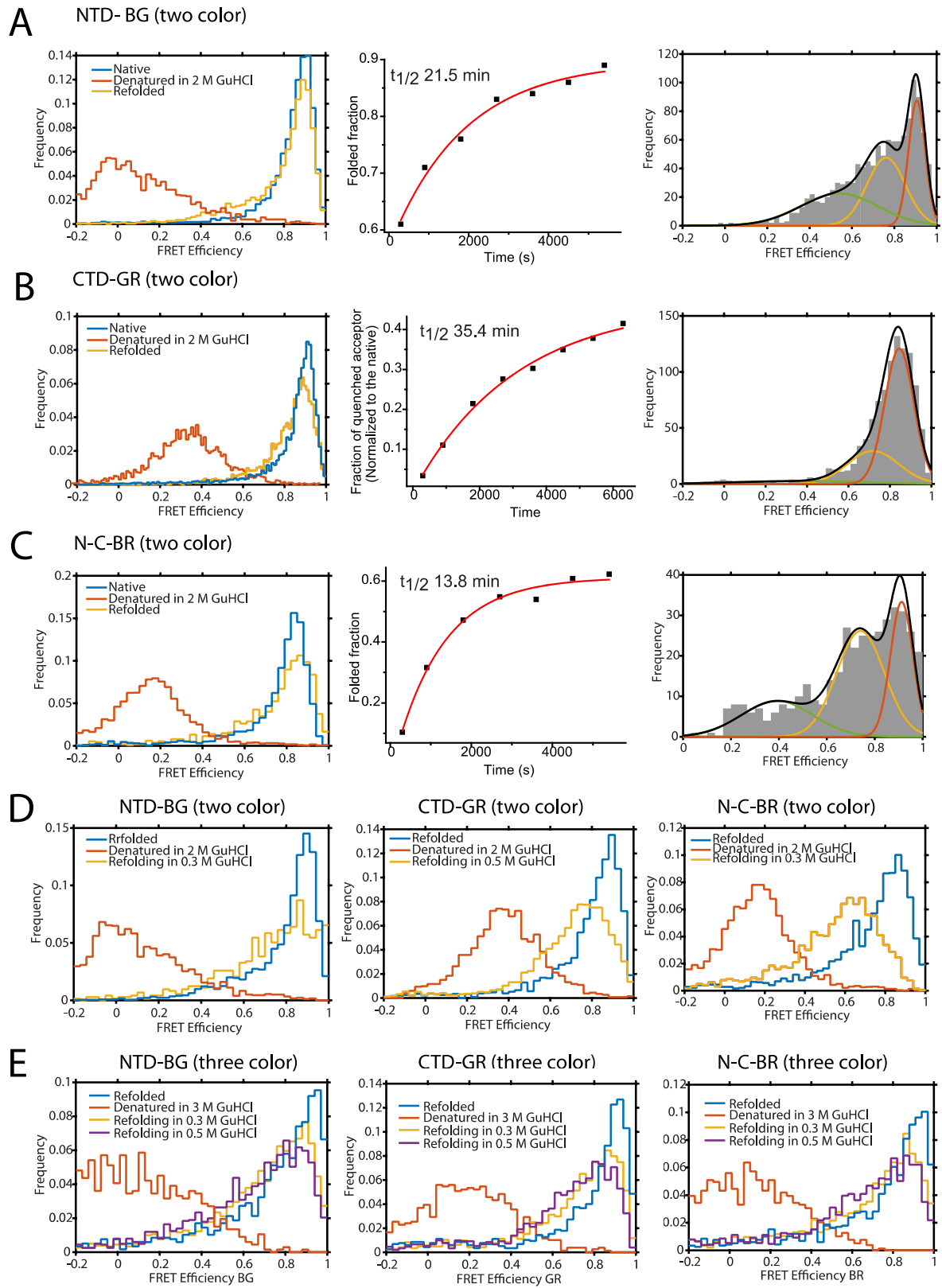

**Supplementary Figure 10. Two-color smFRET control measurements for the three-color smFRET measurements**

**(A)** Two-color BG-FRET control measurements of the NTD. The NTD was labeled with Atto488 and Atto565 as used for three-color smFRET. Left panel: Two-color smFRET histograms comparing the native (blue), 2 M GuHCl denatured (orange) and refolded NTD (yellow). Middle panel: A kinetic analysis of the smFRET histogram during refolding of the NTD. See also Table 1. Right panel: SmFRET histogram for the refolding measurement for which rate was quantified in the middle panel. The unfolded state is shown in green, the intermediate state in yellow and the folded in orange.

**(B)** Two-color GR-FRET control measurements of the CTD. The CTD was labeled with Atto565 and Alexa647 as used for three-color smFRET. Left panel: Two-color smFRET histograms comparing the native (blue), 2 M GuHCl denatured (orange) and refolded CTD (yellow). Middle panel: A kinetic analysis of the smFRET histogram during refolding of the CTD. See also Table 1 and Table S2. Right panel: SmFRET histogram for the refolding measurement for which rate was quantified in the middle panel. The unfolded state is shown in green, the intermediate state in yellow and the folded in orange.

**(C)** Two-color BR-FRET control measurements of the N-C interface. The N-C interface was labeled with Atto488 and Alexa647 as used for three-color smFRET. Left panel: Two-color smFRET histograms comparing the native (blue), 2 M GuHCl denatured (orange) and refolded N-C interface (yellow). Middle panel: A kinetic analysis of the smFRET histogram during refolding of the N-C interface. See also Table 1. Right panel: SmFRET histogram for the refolding measurement for which rate was quantified in the middle panel. The unfolded state is shown in green, the intermediate state in yellow and the folded in orange.

**(D-E)** Comparison for the 1-D smFRET histograms for two-color smFRET control measurements (**D**) and for the three-color smFRET experiments (**E**). The two-color FRET NTD was labeled with same fluorophores (i.e. Atto488 and Atto565) as used for the three-color smFRET measurements. Left panels: BG FRET histograms of the NTD were compared under the conditions of refolded (in 0.1 GuHCl, blue), 2 M GuHCl denatured (orange), and refolding in 0.3 M GuHCl (yellow) and 0.5 M GuHCl (purple). The two-color FRET CTD was labeled with same fluorophores (i.e. Atto565 and Alexa647) as used for the three-color smFRET measurements. Middle panel: GR FRET histograms of the CTD were compared under the conditions of refolded (in 0.1 GuHCl, blue), 2 M GuHCl denatured (orange), and refolding in 0.3 M GuHCl (yellow) and 0.5 M GuHCl (purple). Right panel: BR FRET histograms of N-C interface were compared under the conditions of refolded (in 0.1 GuHCl, blue), 2 M GuHCl denatured (orange), and refolding in 0.3 M GuHCl (yellow) and 0.5 M GuHCl (purple). The two-color FRET N-C interface was labeled with same fluorophores (i.e. Atto488 and Alexa647) as used for the three-color smFRET measurements. **(**

# Supplementary Figure 11

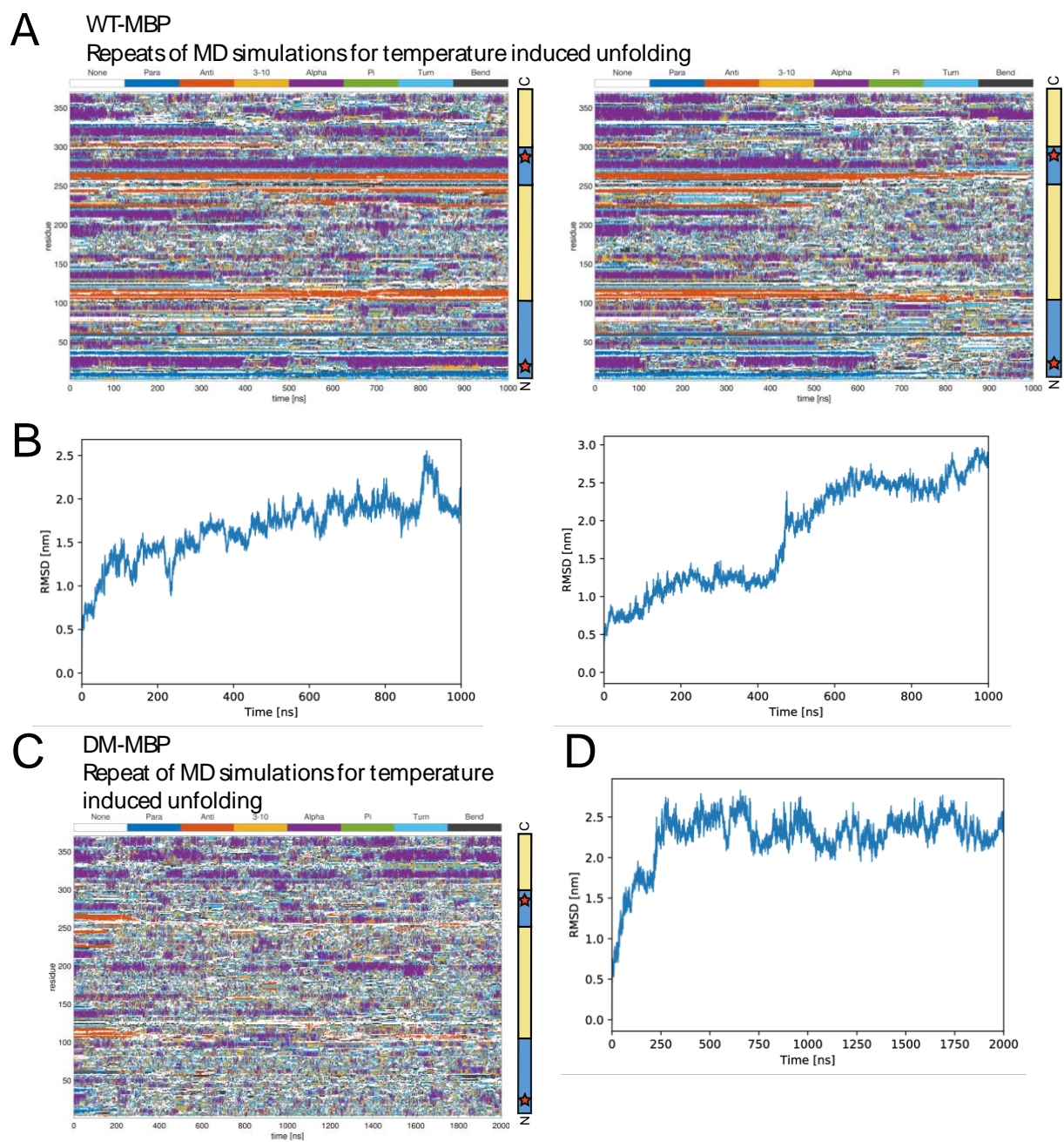

**Supplementary Figure 11. Repeats of MD simulations performed on WT-MBP and DM-MBP for temperature induced unfolding.**

**(A)** DSSP (Definition of Secondary Structure of Proteins) based secondary structure annotation plot for MD simulations performed on WT-MBP (PDB ID:1OMP) for temperature induced unfolding. The plot is obtained for various secondary structures present in amino acid sequence from N to C terminus. Simulations was performed at 450 K for 1 μs. Random coil is shown in white, parallel beta sheet in blue, anti-parallel beta sheet in orange, 3-10 helix in yellow, alpha helix in purple, Pi helix in green, beta turn in sky blue and bends in black.

Predominantly found alpha helices and anti-parallel beta sheets secondary structures can be seen throughout the MBP structure. Double mutations present in DM-MBP, V8G and Y283D are depicted in right panel in MBP sequence. Left and right panels are separate runs starting with the WT-MBP structure (PDB ID:1OMP). As significant changes in the MBP structure were not observed during equilibration step(Figure 7C), we perform the unfolding repeats without first equilibrating the structure.

**(B)** RMSD (Root mean square deviation) plot for MD simulations performed on WT-MBP for temperature induced unfolding shown in panel A. Left and right panel are for respective repeats of the run in panel A. The RMSD is calculated for backbone atoms over the duration of the MD simulation with respect to the initial structure.

**(C)** DSSP based secondary structure annotation plot for MD simulations performed on DM-MBP (PDB ID:1OMP) for temperature induced unfolding. The plot is obtained for various secondary structures present in amino acid sequence from the N to C terminus. The simulation was performed at 450 K for 1 µs. Random coil is shown in white, parallel beta sheet in blue, anti-parallel beta sheet in orange, 3-10 helix in yellow, alpha helix in purple, Pi helix in green, beta turn in sky blue and bends in black. Predominantly found alpha helices and anti-parallel beta sheets secondary structures can be seen throughout the simulation. The double mutations present in DM-MBP, V8G and Y283D, are depicted in the right panel in the MBP sequence.

**(D)** RMSD plot for MD simulations performed on DM-MBP for temperature induced unfolding shown in panel C**.** **The** RMSD is calculated for backbone atoms over the duration of the MD simulation with respect to the initial structure.

## Supplementary Tables

**Table S1:** Fluorescence lifetimes of Alexa647 fit with a single- or double-exponential function for double-labeled molecules and their corresponding fractions ($f$) for the various of DM-MBP constructs. The double cysteine mutants were stochastically labeled with the respective dyes. (See Supplementary Figure 3A).

|  |  | $\tau_1$ (ns) | $f_1$ | $\tau_2$ (ns) | $f_2$ |
|---|---|---|---|---|---|
| N-C interface (52C-175C) Atto532-Alexa647 | Native | 1.34 | 1 | - | - |
|  | Refolding, initial 5 min | - | - | 1.67 | 1 |
|  | Refolded | 1.29 | 0.75 | 1.86 | 0.25 |
|  | Denatured | 1.51 | 1 | - | - |
| CTD (175C-298C) Atto565-Alexa647 | Native | 1.43 | 1 | - | - |
|  | Refolding, initial 5 min | 1.44 | 0.03 | 1.80 | 0.97 |
|  | Refolded | 1.44 | 0.48 | 1.80 | 0.52 |
|  | Denatured | 1.60 | 1 | - | - |

**Table S2:** Fluorescence lifetimes of Alexa647 for triple-labeled molecules in the DM-MBP (52PrK-175C-298C) construct fit with a single- or double-exponential function along with the respective fractions ($f$). This construct was specifically labeled with Atto488 (at position 52), Atto565 (at position175) and Alexa647 (at position 298) as explained in the Material and Methods.

|  |  | $\tau_1$ (ns) | $f_1$ | $\tau_2$ (ns) | $f_2$ |
|---|---|---|---|---|---|
| DM-MBP (52PrK-175C-298C) | Native | 1.33 | 0.77 | 1.74 | 0.23 |
|  | Refolding, initial 15 min | 1.25 | 0.23 | 1.75 | 0.77 |
|  | Refolded | 1.25 | 0.62 | 1.75 | 0.38 |
|  | Denatured | 1.48 | 1 | - | - |

**Table S3**: Steady-state and time-resolved anisotropies of NTD, CTD and N-C interface constructs measured with Atto532 and Alexa647. Steady-state $\langle r \rangle$ and residual anisotropy $r_\infty$ was calculated for double-labeled molecules from the two-color smFRET measurements. Residual anisotropy was calculated by fitting a time-resolved anisotropy data with single exponential model function as a simple approximation.

| DM-MBP | Donor (Atto532) | Acceptor (Alexa647) |
|---|---|---|

| NTD (52C-298C) | | $\langle r \rangle$ | $r_\infty$ | $\langle r \rangle$ | $r_\infty$ |
|---|---|---|---|---|---|
| Native | | 0.249 | 0.059 | 0.196 | 0.121 |
| Refolding in GuHCl (M) | 0.1 | 0.207 | 0.038 | 0.213 | 0.213 |
| | 0.2 | 0.168 | 0.014 | 0.227 | 0.165 |
| | 0.3 | 0.186 | 0.047 | 0.226 | 0.178 |
| | 0.5 | 0.113 | 0.001 | 0.225 | 0.145 |
| | 0.9 | 0.147 | 0.014 | 0.218 | 0.165 |
| | 2 | 0.086 | 0.001 | 0.182 | 0.062 |
| Unfolding in GuHCl (M) | 0.1 | 0.269 | 0.015 | 0.191 | 0.191 |
| | 0.2 | 0.231 | 0.070 | 0.248 | 0.240 |
| | 0.3 | 0.221 | 0.020 | 0.199 | 0.195 |
| | 0.5 | 0.211 | 0.062 | 0.194 | 0.167 |
| | 0.9 | 0.147 | 0.022 | 0.285 | 0.225 |
| | 2 | 0.072 | 0.001 | 0.191 | 0.170 |
| DM-MBP CTD (175C-298C) | | **Donor (Atto532)** | | **Acceptor (Alexa647)** | |
| | | $\langle r \rangle$ | $r_\infty$ | $\langle r \rangle$ | $r_\infty$ |
| Native | | 0.246 | 0.068 | 0.246 | 0.109 |
| Refolding in GuHCl (M) | 0.1 | 0.213 | 0.033 | 0.239 | 0.148 |
| | 0.2 | 0.178 | 0.014 | 0.242 | 0.178 |
| | 0.3 | 0.163 | 0.011 | 0.241 | 0.183 |
| | 0.5 | 0.140 | 0.011 | 0.233 | 0.173 |
| | 0.9 | 0.126 | 0.004 | 0.236 | 0.169 |
| | 2 | 0.122 | 0.003 | 0.238 | 0.163 |
| Unfolding in GuHCl (M) | 0.1 | 0.237 | 0.068 | 0.224 | 0.167 |
| | 0.2 | 0.236 | 0.064 | 0.220 | 0.169 |
| | 0.3 | 0.234 | 0.052 | 0.219 | 0.166 |
| | 0.5 | 0.234 | 0.051 | 0.223 | 0.170 |
| | 0.9 | 0.202 | 0.029 | 0.231 | 0.172 |
| | 2 | 0.122 | 0.010 | 0.204 | 0.121 |
| DM-MBP N-C (52C-175C) | | **Donor (Atto532)** | | **Acceptor (Alexa647)** | |
| | | $\langle r \rangle$ | $r_\infty$ | $\langle r \rangle$ | $r_\infty$ |
| Native | | 0.185 | 0.034 | 0.209 | 0.200 |
| Refolding in GuHCl (M) | 0.1 | 0.193 | 0.015 | 0.239 | 0.188 |
| | 0.2 | 0.198 | 0.045 | 0.241 | 0.217 |
| | 0.3 | 0.197 | 0.047 | 0.258 | 0.201 |

|  | 0.5 | 0.182 | 0.033 | 0.230 | 0.211 |
|---|---|---|---|---|---|
|  | 0.9 | 0.143 | 0.021 | 0.229 | 0.200 |
|  | 2 | 0.127 | 0.012 | 0.235 | 0.115 |
| Unfolding in GuHCl (M) | 0.1 | 0.203 | 0.051 | 0.193 | 0.190 |
|  | 0.2 | 0.200 | 0.040 | 0.198 | 0.189 |
|  | 0.3 | 0.195 | 0.040 | 0.208 | 0.199 |
|  | 0.5 | 0.200 | 0.033 | 0.202 | 0.201 |
|  | 0.9 | 0.160 | 0.016 | 0.212 | 0.190 |
|  | 2 | 0.160 | 0.016 | 0.203 | 0.148 |